\documentclass{aa}  

\usepackage{graphicx}
\usepackage{txfonts}
\usepackage{epsfig}
\usepackage{amsmath}
\usepackage{amssymb}
\usepackage{makecell}
\usepackage{natbib}
\usepackage{color} 
\begin{document} 

%%%%%%%%%%%%%%%%%%%%%%%%%%%%%%%%%%%%%%%%%%%%%%%%%%

%%%%% AUTHORS - PLACE YOUR OWN COMMANDS HERE %%%%%

% Please keep new commands to a minimum, and use \newcommand not \def to avoid
% overwriting existing commands. Example:
%\newcommand{\pcm}{\,cm$^{-2}$}	% per cm-squared

\def\aap{A\& A}
\def\AJ{{AJ}}
\def\aj{{AJ}}
\def\apj{ApJ}
\def\apjs{ApJS}
\def\apjl{ApJL}
\def\apss{Ap\& SS}
\def\ARAA{{ARA\&A}}
\def\araa{ARA\&A}
\def\nat{Nature}
\def\mnras{{MNRAS}}
\def\MNRAS{MNRAS}
\def\Nat{Nature}
\def\Obs{{ Observatory}}
\def\pasj{{PASJ}}
\def\pasp{{PASP}}
\def\rmaa{{RevMexAA}}

%%%%%%%%%%%%%%%%%%%%%%%%%%%%%%%%%%%%%%%%%%%%%%%%%%%%%%%%%%%%%%%%%%%%%%%%%
   \title{Extreme fragmentation and complex kinematics at the center of the L1287 cloud}
   \titlerunning{L1287}
%   \subtitle{I. Overviewing the $\kappa$-mechanism}

   \author{Carmen Ju\'arez
          \inst{1,2,3}, 
          Hauyu Baobab Liu
          \inst{4}, 
          Josep M. Girart
          \inst{1,2} 
          \and
          Aina Palau
          \inst{5}
          \and
          Gemma Busquet
          \inst{1,2},
          Roberto Galv\'an-Madrid
          \inst{5},
          Naomi Hirano
          \inst{6}
          \and
          Yuxin Lin
          \inst{7,8}
          %Josep M. Girart\inst{1,3}\fnmsep\thanks{Just to show the usage
          %of the elements in the author field}
          }
	\authorrunning{Ju\'arez et al.}
   \institute{$^1$Institut de Ci\`encies de l'Espai, CSIC,
              Campus UAB, Carrer de Can Magrans, S/N, 08193 Cerdanyola del Vall\`es, Barcelona, Catalonia, Spain\\
              \email{carmenjr27@gmail.com}\\
              $^2$Institut d'Estudis Espacials de Catalunya (IEEC), Barcelona, Catalonia, Spain\\
              $^3$Dept. de F\'{\i}sica Qu\`antica i Astrof\'{\i}sica, Institut de Ci\`encies del Cosmos (ICCUB)\thanks{The ICCUB is a CSIC-Associated Unit through the Institut de Ci\`encies de l'Espai (ICE).}, Universitat de Barcelona (IEEC-UB), Mart\'{\i} Franqu\`es 1, E08028 Barcelona, Catalonia, Spain\\
              %\email{juarez@ice.cat}
             %        accept e-mails}
              $^4$European Southern Observatory (ESO), 
              Karl-Schwarzschild-Str. 2, D-85748 Garching, Germany\\
              %\email{juarez@ice.cat}
              $^5$Instituto de Radioastronom\'ia y Astrof\'isica,
              Universidad Nacional Aut\'onoma de M\'exico, P.O. Box 3-72, 58090, Morelia, Michoac\'an, Mexico\\
              %\email{juarez@ice.cat}
              $^6$Institute of Astronomy and Astrophysics, Academia Sinica,
              P.O. Box 23-141, Taipei 106, Taiwan\\
              $^7$National Astronomical Observatories, Chinese Academy of Sciences\\
              $^8$Max-Planck-Institut f\"{u}r Radioastronomie, D-53121 Bonn, Germany\\
             %\email{c.ptolemy@hipparch.uheaven.space}
             %\thanks{The university of heaven temporarily does not
             %        accept e-mails}
             }

   %\date{Received September 15, 1996; accepted March 16, 1997}

% \abstract{}{}{}{}{} 
% 5 {} token are mandatory

  \abstract
  {} % leave it empty if necessary  
%   {L1287 is a dark cloud located at 929~pc. It harbors a binary FU-Orionis system, RNO1B/1C, and several centimeter and infrared sources, including IRAS 0038+6312. Previous observations have found a bipolar outflow in a northeast-southwest direction and several sources have been proposed as the powering source.}
   {
   The filamentary, $\sim$10 pc scale infrared dark cloud L1287 located at a $\sim$929 pc parallax distance, is actively forming a dense cluster of low-mass young stellar objects (YSOs) at its inner $\sim$0.1 pc region. To help understand the origin of this low-mass YSO cluster, the present work aims at resolving the gas structures and kinematics with high angular resolution. 
   %We aim to study the dust and gas properties of L1287 to analyze the dynamical evolution of this low-mass cluster-forming region. In addition we try to provide a more definite identification of the powering source of the previously reported outflow.
   }
   {We have performed $\sim$1$''$ angular resolution ($\sim$930 AU) observations at $\sim$1.3~mm wavelengths using the Submillimeter Array (SMA), which simultaneously covered the dust continuum emission, and various molecular line tracers for dense gas, warm gas, shocks, and outflows.}
   {From a $\sim$2$''$ resolution 1.3~mm continuum image we identified six dense cores, namely SMA1-6. Their gas masses are in the range of $\sim0.4-4$~M$_\odot$. From a $\sim$1$''$ resolution 1.3~mm continuum image, we find a high fragmentation level, with 14 compact millimeter sources within 0.1 pc: SMA3 contains at least nine internal condensations; SMA5 and SMA6 are also resolved with two internal condensations. Intriguingly, one condensation in SMA3, and one condensation in SMA5, appear associated with the known accretion outburst YSOs RNO\,1C and RNO\,1B.
The dense gas tracer DCN (3--2) traces well the dust continuum emission and shows a clear velocity gradient along the NW-SE 
direction centered at SMA3. There is another velocity gradient with opposite direction around the most luminous young stellar object IRAS 00338+6312.}
   {%The dynamics in the region can be understood by a single scenario of converging flows+rotation towards IRAS 0038+6312. In addition, IRAS 0038+6312 seems to be the most likely powering source of the previously reported bipolar outflow. FU-Orionis RNO1C seems to be the powering source of an additional molecular outflow.
  {The fragmentation within 0.1 pc in L1287 is very high compared to other regions at the same spatial scales. The incoherent motions of dense gas flows are sometimes interpreted by being influenced by (proto)stellar feedback (e.g., outflows), which is not yet ruled out in this particular target source. On the other hand, the velocities (with respect to the systemic velocity) traced by DCN are small, and the directions of the velocity gradients traced by DCN are approximately perpendicular to those of the dominant CO outflow(s). Therefore, we alternatively  hypothesize that the velocity gradients revealed by DCN trace the convergence from the $\gtrsim$0.1~pc scales infalling motion towards the rotational motions around the more compact ($\sim0.02$~pc) sources. This global molecular gas converging flow may feed the formation of the dense low-mass YSO cluster. Finally, we also found that IRAS 00338+6312 is the most likely powering source of the dominant CO outflow. 
A compact blue-shifted outflow from RNO\,1C is also identified.
  }
  }

   \keywords{Stars: formation -- 
   		ISM: individual objects: L1287 --
        ISM: molecules 
               }

   \maketitle
%
%-------------------------------------------------------------------

%%%%%%%%%%%%%%%%%%%%%%%%%%%%%%%%%%%%%%%%%%%%%%%%%%%%%%%%%%%%%%%%%%%%%%%%%%%%%%%%%%%%%%%%%%%%%%%%%%%%%%%%%%%%
\section{Introduction}\label{sec:introduction}

Molecular gas filaments collapsing towards the center of molecular clouds may pile up where the centrifugal force is becoming important as compared with the gravitational force, and then form dense stellar clusters.
The observationally very well resolved examples include the $L\sim$10$^{7}$~$L_{\odot}$ OB cluster-forming region W49A \citep[e.g.,][]{Keto1991,Galvan2013,Lin2016}, the $L\sim$10$^{6}$ $L_{\odot}$ OB cluster-forming region G10.6-0.4 \citep[e.g.,][]{Keto1987,Liu2012a}, the $L\sim$10$^{5}$~$L_{\odot}$ OB cluster-forming region G33.92+0.11 \citep[e.g.,][]{Liu2012b,Liu15}, and the $L\sim$10$^{4}$ $L_{\odot}$ OB cluster-forming region NGC6334\,V \citep[e.g.,][]{Juarez17}.
Is it plausible to form dense clusters of low-mass young stellar objects (YSOs) with a similar mechanism \citep[c.f.,][]{Corsaro2017,Mapelli2017}?

\begin{figure*}
%\vspace{-2cm}
\hspace{-0.4cm}
\begin{tabular}{ p{8.5cm} p{8.5cm} }
\includegraphics[width=9.6cm]{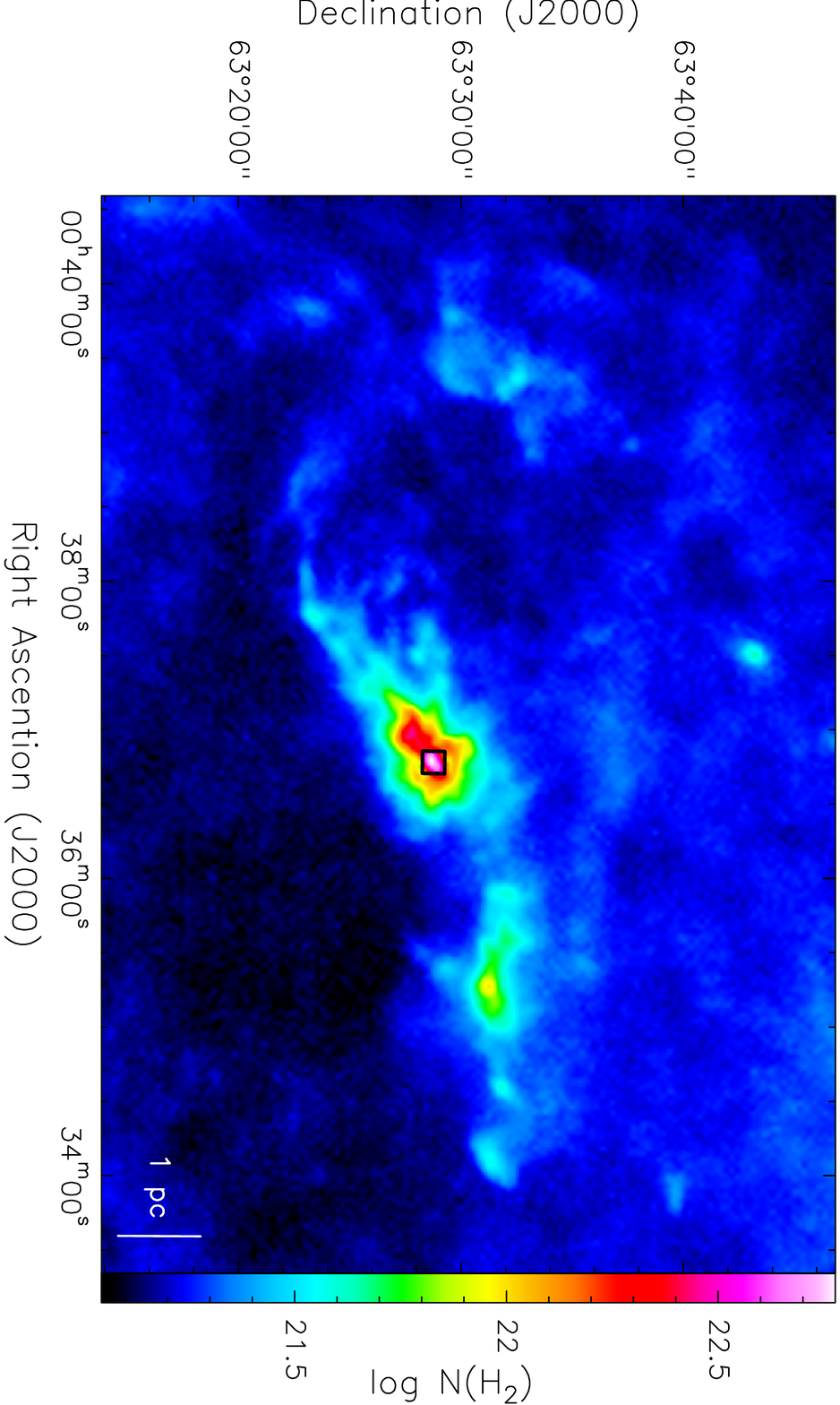} & \hspace{0.8cm}\vspace{-7cm} \includegraphics[width=9.2cm]{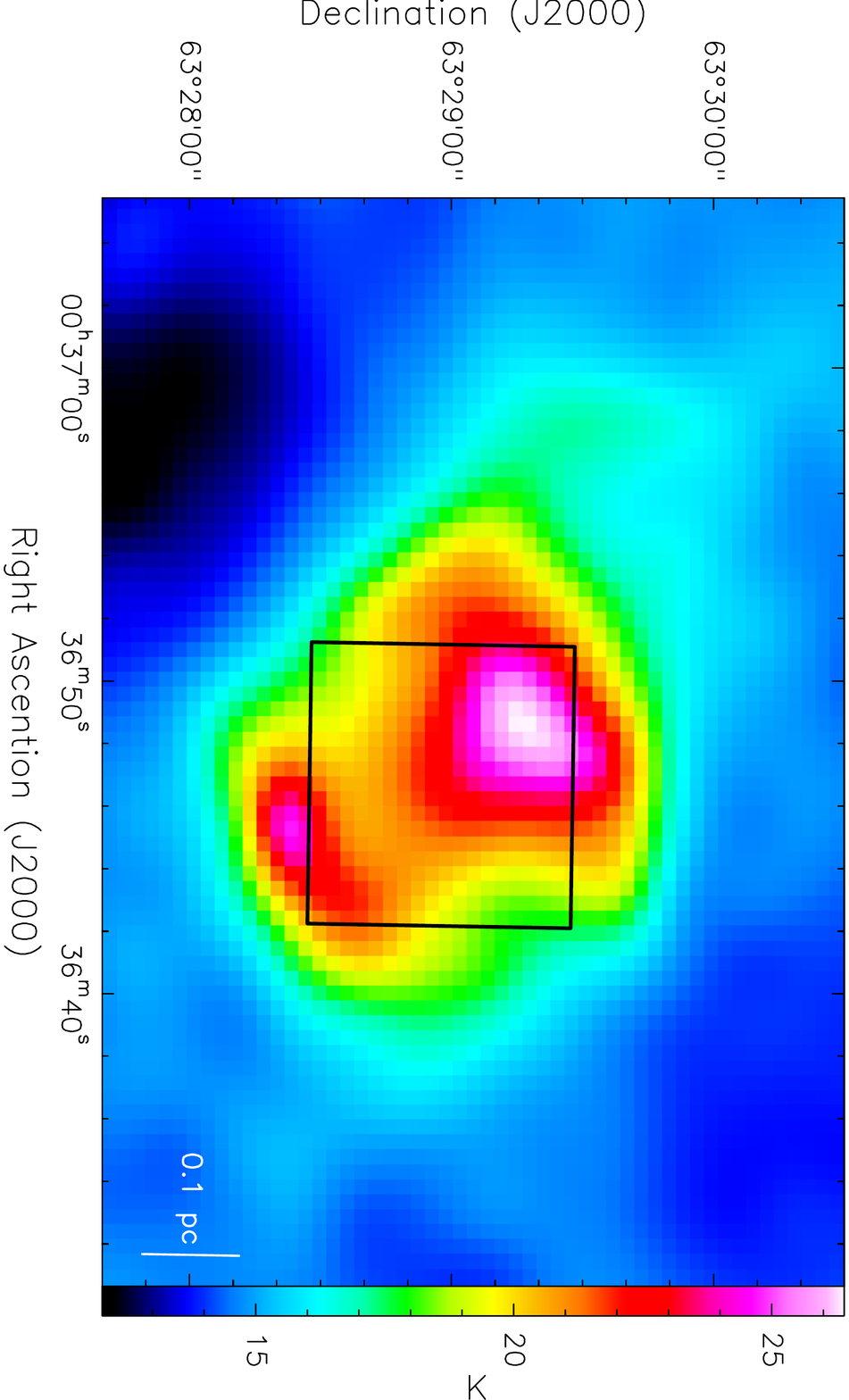} \\
\end{tabular}
\caption{The 37$''$ angular resolution gas column density ($N$(H$_{2}$); {\it left}) and dust temperature ($T_d$; {\it right}) maps of L1287 (for more details see Section \ref{sub:Herschelobs}). 
The black box in both panels indicates the field of view of the top panel of Fig.~\ref{continuum}, which presents the 1.3~mm continuum image taken with the SMA.}
\label{Herschel-N_T}
\end{figure*}

To follow up this issue, we select to study the filamentary dark molecular cloud, L1287, which is located at a distance of $929^{+34}_{-33}$ pc \citep{Rygl10}.
Its inner, densest $\sim$0.1 pc region is forming a $L\sim$10$^{3}$ $L_{\odot}$ low-mass YSO cluster \citep[e.g.][]{Estalella93, Quanz07}.
Four of the YSOs, namely VLA1-4, are known to be associated with thermal radio jets \citep{Anglada94}.
Based on the previous, lower angular resolution observations of C$^{18}$O (1--0), \citet{Umemoto1999} estimated the enclosed gas mass over the $\sim$0.5 pc scale parent molecular gas structure to be $\sim$120 $M_{\odot}$.
% Large-scale images at 160, 250, 350 and 500~$\mu$m from {\it Herschel} \citep{Pilbratt10}, tracing cold dust towards the region, show a $\sim$10~pc projected scale filamentary structure with L1287 located approximately at the center (see Fig.~\ref{fig:herschel}).
%
Contrasting to the aforementioned OB cluster-forming regions, the bolometric luminosity of L1287 is not contributed by the nuclear burning of OB stars, but instead, is contributed by the accretion luminosity of the low-mass protostars, including two known FU Orionis objects\footnote{FU Orionis objects are young, pre-main-sequence stars which are observed to increase their brightness by 4--6 mag in the optical and remain bright for decades \citep{Herbig77}. The large-amplitude flares are attributed to enhanced accretion from the surrounding circumstellar disk \citep{Hartmann&Kenyon85}.} RNO\,1B/1C \citep{Staude91,Kenyon93,McMuldroch95,Quanz07}, and a $L\sim600$ L$_\odot$ Class\,0/I YSO IRAS\,00338+6312 \citep{Anglada94}.
Earlier observations have resolved bipolar CO outflow(s) aligning in the northeast-southwest direction \citep{Snell90,Yang91,Feher2017}, although the identification of the powering source(s) remains uncertain. % \citep[e.g., RNO\,1B/1C by][]{Staude91,McMuldroch95}, \citep[e.g., VLA\,3 by][]{Anglada94}, and  \citep[e.g., IRAS\,0038+6312 by][]{Yang91,Quanz07}.

% RNO1B/1C belong to a young, small stellar cluster. These young stellar objects (YSO) are detected in the near-IR and cm wavelengths \citep[][see Fig.~\ref{continuum}]{Anglada94,Quanz07}. 
% All infrared sources (except for two) have been classified as Class 0/I-II objects, including RNO1B/1C \citep{Quanz07}. 
% The region also harbors a very deeply embedded protostar, IRAS 00338+6312. 
% The bolometric IR luminosity of IRAS 00338+6312 calculated from the fluxes of the four IRAS bands is $\sim600$ L$_\odot$. \citet{Anglada94} proposed that their cm source VLA~3 is associated with IRAS 00338+6312.

% We aim to study the dust and gas properties of L1287 to analyze the dynamical evolution of this low-mass cluster-forming region. In addition we try to provide a more definite identification of the powering source of the previously reported outflow. 

\begin{table*}
\caption{SMA observations summary.}
\begin{center}
	\begin{tabular}{lcccc}
	\hline
	\hline
	Array configuration			&subcompact				&compact		&extended			\\\hline
	Observing date				&2014 Feb 10, July 25	&2013 Aug 2	&2013 Oct 24, 25		\\
	$\tau_{225}$$_\text{GHz}$	&0.14, 0.12				&0.13		&0.25, 0.16				\\
	Number of antennas			&6, 7					&5			&6						\\
	Time on target (hours)		&2, 5					&8			&8						\\
	Flux calibrator				&Callisto, Neptune		&Uranus		&Uranus					\\
	Passband calibrator			&0102+584				&3c84		&3c454.3				\\
	Gain calibrator				&0102+584				&0102+584 	&0102+584				\\ \hline
	\end{tabular}
\end{center}
\label{observations}
\end{table*}

In this paper, we report the $\sim$1$''$ ($\sim$930 AU) angular resolution observations with the Submillimeter Array (SMA)\footnote{The Submillimeter Array is a joint project between the Smithsonian
Astrophysical Observatory and the Academia Sinica Institute of Astronomy
and Astrophysics, and is funded by the Smithsonian Institution and the
Academia Sinica \citep{Ho04}.}, covering the inner $\sim$0.25 pc region of L1287.
By examining the resolved 1.3~mm dust continuum emission and various molecular line tracers of dense gas, warm gas, shocks, and outflows, our aim is to address how the gas flows converge from parsec scales (e.g., due to gravitational collapse), down to $\sim1000$~AU scales, and is dynamically forming the low-mass YSO cluster.
We also provide a more definite identification of the powering source of the previously reported molecular outflows.

The details of our observations are presented in Section~\ref{sec:observations}.
The observational results from the dust continuum and molecular line emission are given in Section~\ref{sec:results}. 
In Section~\ref{sec:analysis} we present an analysis of fragmentation, dense gas kinematics, and powering sources of the previously reported CO outflows. 
Our conclusions are given in Section~\ref{sec:summary}.

%%%%%%%%%%%%%%%%%%%%%%%%%%%%%%%%%%%%%%%%%%%%%%%%%%%%%%%%%%%%%%%%%%%%%%%%%%%%%%%%%%%%%%%%%%%%%%%%%%%%%%%%%%%%
%%%%%%%%%%%%%%%%%%%%%%%%%%%%%%%%%%%%%%%%%%%%%%%%%%%%%%%%%%%%%%%%%%%%%%%%%%%%%%%%%%%%%%%%%%%%%%%%%%%%%%%%%%%%

\section{Observations and data reduction} \label{sec:observations}

\subsection{Submillimeter Array observations}\label{sub:SMAobs}
We have performed five tracks of SMA observations at $\sim$1.3~mm wavelengths towards the inner $\sim$0.25 pc region of L1287, in between August 2013 and July 2014, which are summarized in Table~\ref{observations}.
The pointing and phase referencing centers of the observations are RA(J2000)$=00^{\rm h} 36^{\rm m} 46.65^{\rm s}$, Dec(J2000)$=+63^{\circ}28'57.90''$.
The projected baselines of these observations range from $\sim$4~k$\lambda$ to $\sim$140~k$\lambda$.
The system temperature T$_\text{sys}$ for all the observations was around 150 K.

The observations were obtained with the 230~GHz receiver, with 4~GHz bandwidth in each of the upper and lower sidebands.
The correlator consisted of 48 chunks with a bandwidth of 104~MHz each.
The frequency was centered at 231.3~GHz at the chunk 35 of the upper sideband (USB). 
This configured the correlator with the lower sideband (LSB) covering from 216.48~GHz to 220.46~GHz and the USB from 228.46~GHz to 232.43~GHz.
These frequency ranges covered the molecular lines listed in Table~\ref{transitions}. 
The $^{13}$CO and N$_2$D$^+$ lines were observed with 512 channels per chunk and a spectral resolution of 0.26~km~s$^{-1}$, the line NH$_2$D was observed with 256 channels per chunk giving 0.53 km s$^{-1}$ spectral resolution. 
The rest of the lines were observed with 128 channels per chunk and a 1.06~km~s$^{-1}$ spectral resolution. 
To generate the 1.3~mm continuum, we averaged the line-free channels in the lower and upper sidebands.
The continuum data have been reported by \citet{Liu2018} as part of a SMA survey towards 29 accretion outburst YSOs, which did not analyze the details of the extended structures, and did not analyze spectral line data.
 
  %These frequency ranges include the following molecular lines: CO (2--1), $^{13}$CO (2--1), SO (6(5)--5(4)), SiO (5--4), C$^{18}$O (2--1), N$_2$D$^+$ (3--2) and NH$_2$D (3(2,2)--3(1,2)).

% The weather conditions in all the observing dates were moderately good. 
 
The calibration for absolute flux, bandpass, and gain were carried out using the {\sc mir idl} software package\footnote{https://www.cfa.harvard.edu/$\sim$cqi/mircook.html}.
The images were created using the Multichannel Image Reconstruction, Image Analysis, and Display \citep[{\sc Miriad},][]{Sault95} software package.

To map the rather extended gas flows, we used {\sc Robust}$=1$ weighting \citep{Briggs95} to create the 1.3~mm dust continuum emission map using all data listed in Table~\ref{observations}, which yielded a $2\farcs28\times2\farcs17$ (P.A. $=-19^\circ$) synthesized beam, and a root-mean square (RMS) noise level is 0.6 mJy beam$^{-1}$. 
In addition, we created a higher angular resolution 1.3~mm dust continuum emission map of spatially compact sources, using only the observations taken in the extended array configuration. 
We used a {\sc Robust}$=0$ weighting, which yielded a synthesized beam of $0\farcs96\times0\farcs79$ (P.A.$= -22^\circ$), and an RMS noise level of 0.5 mJy beam$^{-1}$.

%We used {\sc Robust} 0 \citep{Briggs95} to create the 1.3~mm dust continuum emission map with only the extended configuration. 

To present the maps of the molecular lines, we applied three different weightings. For most of the lines that show extended emission we used a {\sc Robust}$=2$ weighting (i.e, natural weighting), which provides the best sensitivity. 
For the CO (2--1) line we used a {\sc Robust}$=0$ weighting to obtain a better angular resolution. 
Finally, for OCS (18--17), HNCO 10(0,10)--9(0,9) and CH$_3$OH 3($-2$,2)--4($-1$,4), which present spatially compact emission, we used only data from the extended and compact array configurations and a {\sc Robust}$=2$ weighting.

\subsection{Archival Herschel images}\label{sub:Herschelobs}
To present the structures of the parent molecular cloud on 1-10 pc scales, and to estimate the dust temperature, we retrieved the Herschel-SPIRE 250, 350, and 500~$\mu$m image, which were taken as part of the Herschel Infrared Galactic Plane (Hi-GAL) survey \citep[obsID:1342249229,][]{Molinari10}.

We smoothed the 250 and 350~$\mu$m images to the angular resolution of the 500~$\mu$m image (37$''$), and then derived gas column density ($N$(H$_{2}$)) and dust temperature ($T_d$) pixel-by-pixel by fitting the modified-black-body spectrum
\begin{equation}
S_{\nu} = \Omega_{m}B_{\nu}(T_{d})(1-e^{-\tau_{\nu}}),
\end{equation}
where $S_{\nu}$ is the flux observed at frequency $\nu$, $\Omega_{m}$ is the considered solid angle, $B_{\nu}(T_d)$ is the Planck function at $T_{d}$, and $\tau_\nu$ is the dust optical depth. 
We converted $\tau_\nu$ to the gas column density $N(H_{2})$ by assuming
\begin{equation}
N(H_{2}) = R\frac{\tau_{\nu}}{\kappa_{\nu}\mu m_{H}},
\end{equation}
where $R$ is the gas-to-dust mass ratio which we assumed to be 100, $\mu$ = 2.8 is the mean molecular weight, m$_{H}$ is the mass of a hydrogen atom, and $\kappa_{\nu}$ is the dust mass opacity which is assumed to have the following dependence on frequency $\nu$ 
\begin{equation}
\kappa_{\nu} = \kappa_{\mbox{\scriptsize 230 GHz}}\,\Bigg[\frac{\nu \mbox{\,[GHz]}}{230}\Bigg]^{\beta},
\end{equation}
where $\kappa_{230}$ = 0.9~cm$^{2}$g$^{-1}$ \citep{Ossenkopf94}. We assumed a dust emissivity index $\beta$ of 1.8.
We refer to \citet{Hildebrand1983} for an introduction to the formulation we are based on.

%
%%%%%%%%%%%%%%%%%%%%%%%%%%%%%%%%%%%%%%%%%%%%%%%%%%%%%%%%%%%%%%%%%%%%%%%%%%%%%%%%%%%%%%%%%%%%%%%%%%%%%

%
\begin{figure}
%\vspace{-2cm}
\hspace{-0.4cm}
\begin{tabular}{ p{8.5cm} p{8.5cm} }
\includegraphics[width=9cm]{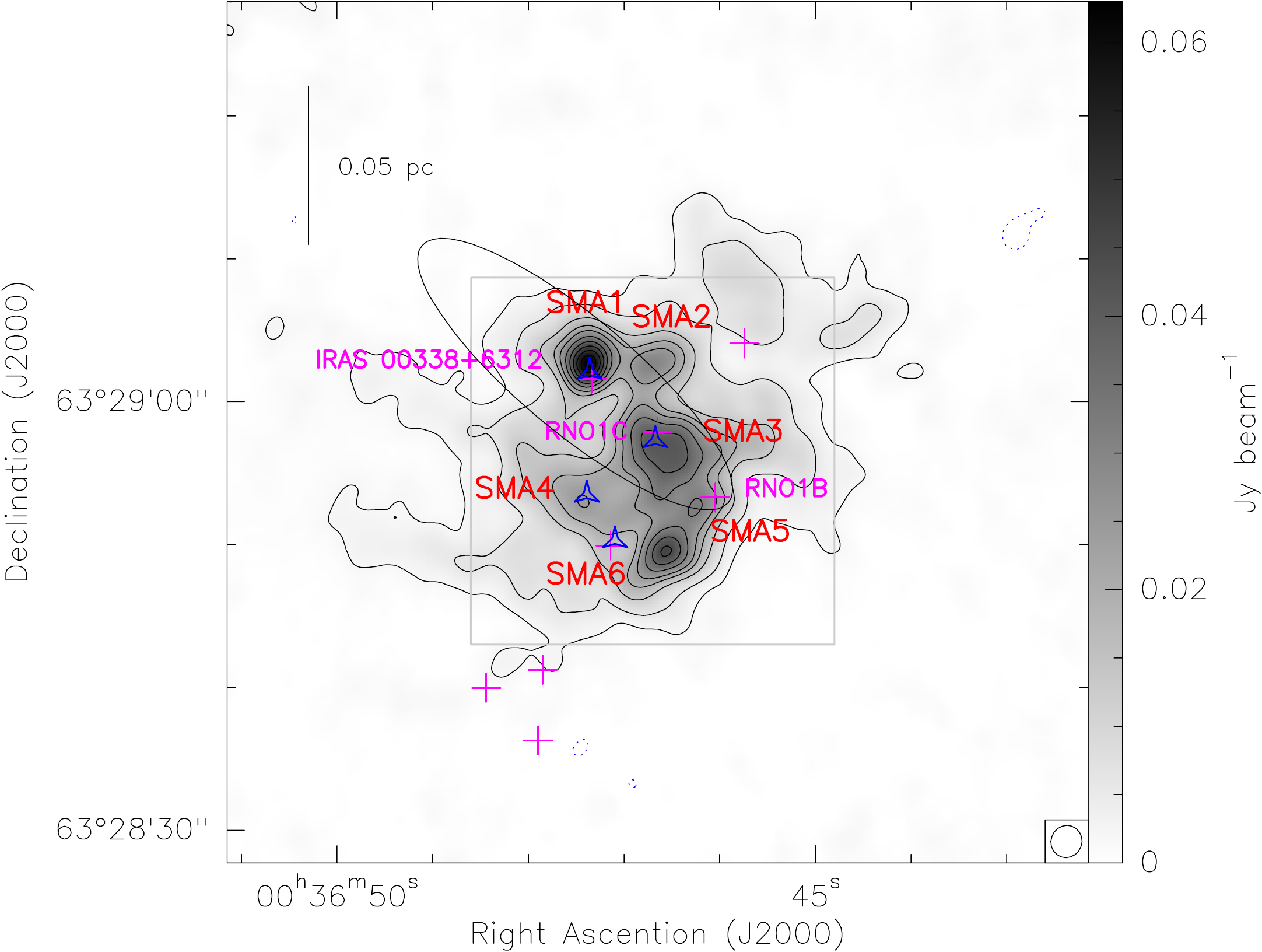} \\ 
\hspace{0.1cm}\vspace{-7cm} 
\includegraphics[width=9cm]{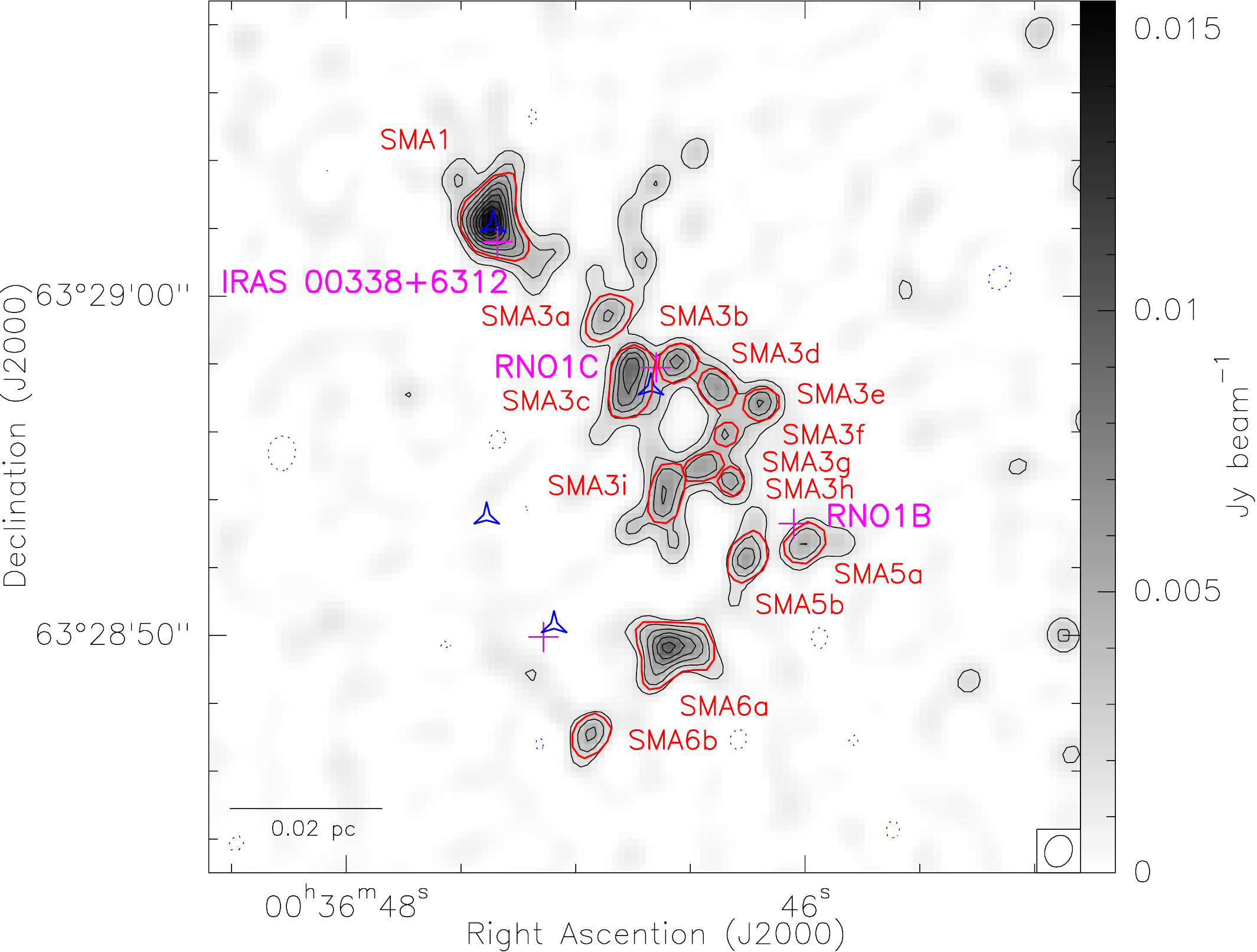} \\
\end{tabular}
\caption{{\it Top}: 1.3 mm dust continuum emission with extended, compact and subcompact configurations. Contours are $-6$, $-3$, 3, 6, 9, 20, 30, 38, 49, 60, 70, 80, 90 
times the rms level of the map, 0.6 mJy beam$^{-1}$. Black ellipse corresponds to the position error ellipse for IRAS 00338+6312. The synthesized beam located at bottom right is $2\farcs28\times2\farcs17$, P.A. $=-19^\circ$. Gray square corresponds to the field of view of the bottom panel. A scale bar is located at the upper left corner. 
{\it Bottom}: 1.3 mm dust continuum emission with extended configuration. Contours are $-3$, 3, 6, 9, 12, ..., 30  times the rms level of the map, 0.5 mJy beam$^{-1}$. Sources (labeled in red) are defined having at least a 7 $\sigma$ closed contour. Red contours are drawn following the 6 and 9 sigma contour levels for the most extended sources. The synthesized beam located at bottom right is $0\farcs96\times0\farcs79$, P.A. $=-22^\circ$. A scale bar is located at the lower left corner. Blue concave hexagons and pink crosses in both panels are radio \citep{Anglada94} and infrared \citep{Quanz07} sources respectively.}
\label{continuum}
\end{figure}
%

%----------------------------------------------------------------------------
\begin{table*}
\caption{Parameters of the sources detected with the SMA at 1.3~mm dust continuum emission.}
\begin{center}
{\small
\begin{tabular}{lccccccccc}
\noalign{\smallskip}
\hline\noalign{\smallskip}
&\multicolumn{2}{c}{Position$^\mathrm{a}$}
&Deconv.ang.size$^\mathrm{a}$
&P.A.$^\mathrm{a}$
&$I_\mathrm{\nu}^\mathrm{peak}$~$^\mathrm{a}$
&$S_\mathrm{\nu}$~$^\mathrm{b}$
&Mass$^\mathrm{c}$
\\
\cline{2-3}
Source
&$\alpha (\rm J2000)$
&$\delta (\rm J2000)$
&$(''\times'')$
&($^\circ$)
%&(m\jpb)
&(mJy beam$^{-1}$)
&(mJy)
%&(\mo)
&(M$_\odot$)
&Association$^\mathrm{d}$\\
\noalign{\smallskip}
\hline\noalign{\smallskip}
SMA1	 	&00:36:47.366 &63:29:02.237		&$1.6\pm0.3\times0.9\pm0.2$  	&$8\pm15$ 	&$13.84\pm1.72$	&$41.4\pm8.3$	&0.69 &CM:VLA~3/IR:IRAS\\

SMA3a 		&00:36:46.860 &63:28:59.414		&$1.2\pm0.2\times0.6\pm0.1$  	&$-44\pm8$ 	&$4.71\pm0.28$ 	&$9.2\pm1.9$	&0.15 &--\\
SMA3b		&00:36:46.562 &63:28:58.040		&$0.9\pm0.3\times0.6\pm0.2$  	&$-69\pm31$ &$6.45\pm0.66$ 	&$11.1\pm2.3$ 	&0.19 &IR:RNO1C\\
SMA3c  		&00:36:46.761 &63:28:57.583		&$1.9\pm0.4\times0.6\pm0.1$  	&$-9\pm6$ 	&$8.72\pm0.85$ 	&$24.7\pm5.0$	&0.41 &CM:VLA~1\\
SMA3d 		&00:36:46.387 &63:28:57.326		&point source 					&--			&$5.93\pm0.24$ 	&$11.0\pm2.2$	&0.18 &--\\
SMA3e 		&00:36:46.198 &63:28:56.829		&$0.6\pm0.1\times0.5\pm0.1$ 	&$-77\pm29$	&$6.24\pm0.35$ 	&$8.6\pm1.8$ 	&0.14 &--\\
SMA3f 		&00:36:46.348 &63:28:55.930		&$1.0\pm0.3\times0.6\pm0.1$		&$-25\pm11$	&$4.74\pm0.10$ 	&$8.6\pm1.8$ 	&0.14 &--\\
SMA3g		&00:36:46.450 &63:28:54.966		&$1.5\pm0.3\times0.4\pm0.2$ 	&$-70\pm10$ &$5.72\pm0.36$  &$13.0\pm2.7$	&0.22 &--\\	
SMA3h 		&00:36:46.330 &63:28:54.583		&point source					&--			&$5.41\pm0.39$ 	&$7.3\pm1.5$ 	&0.12 &--\\
SMA3i		&00:36:46.606 &63:28:54.248		&$2.0\pm0.6\times0.6\pm0.2$ 	&$-13\pm8$ 	&$6.22\pm0.58$  &$18.4\pm3.8$	&0.31 &--\\	

SMA5a		&00:36:46.000 &63:28:52.704		&$1.1\pm0.3\times0.5\pm0.2$ 	&$-64\pm17$	&$4.50\pm0.42$ 	&$8.3\pm1.7$ 	&0.14 &IR:RNO1B\\ 
SMA5b 		&00:36:46.255 &63:28:52.307		&$0.9\pm0.2\times0.1\pm0.1$ 	&$-16\pm8$	&$5.63\pm0.46$ 	&$7.6\pm1.6$ 	&0.13 &--\\

SMA6a 		&00:36:46.584 &63:28:49.613		&$1.5\pm0.3\times0.9\pm0.2$ 	&$-82\pm22$	&$8.50\pm0.11$ 	&$25.7\pm5.2$	&0.43 &--\\
SMA6b		&00:36:46.936 &63:28:47.078		&$0.7\pm0.1\times0.2\pm0.1$  	&$-28\pm8$ 	&$4.96\pm0.24$ 	&$6.5\pm1.3$ 	&0.11 &--\\

\hline
\end{tabular}
\begin{list}{}{}
%\item[$^\mathrm{a}$] Rms noise of the map measured using go noise task in mapping.
\item[$^\mathrm{a}$] Position, deconvolved size, position angle (P.A.), peak intensity, flux density and the respective uncertainties derived from fitting a 2D Gaussian to each source using {\sc imfit} task from {\sc Miriad}. %Peak intensities and flux densities (with uncertainties) are corrected for the primary beam response.
\item[$^\mathrm{b}$] Error in flux density has been calculated as $\sqrt{(\sigma\,\theta_\mathrm{source}/\theta_\mathrm{beam})^2+(\sigma_\mathrm{flux-scale})^2}$ \citep{Beltran01,Palau13}, where $\sigma$ is the rms of the map, $\theta_\mathrm{source}$ and $\theta_\mathrm{beam}$ are the size of the source and the beam, respectively, and $\sigma_\mathrm{flux-scale}$ is the error in the flux scale, which takes into account the uncertainty on the calibration applied to the flux density of the source ($S_\nu\times\%_\mathrm{uncertainty}$) which we assumed to be 20\%.
\item[$^\mathrm{c}$] Masses derived assuming a dust temperature of 22~K, and a dust (+gas) mass opacity coefficient at 1.3~mm of 0.9~cm$^2$\,g$^{-1}$ \citep[for thin ice mantles after $10^5$ years of coagulation at a gas density of $10^6$ cm$^{-3}$,][]{Ossenkopf94}. The uncertainty in the masses due to the opacity law and temperature is estimated to be a factor of 4.
\item[$^\mathrm{d}$] Association with signposts of stellar activity: IR = infrared source; CM = Centimeter radio source. IRAS refers to IRAS 0038+6312.
%\item[$^\mathrm{f}$]  Intensity and flux density derived from a Gaussian fit (MM6).
\end{list}
}
\end{center}
\label{parameters}
\end{table*}

\begin{figure*}
\centering
\includegraphics[scale=0.7,keepaspectratio=true]{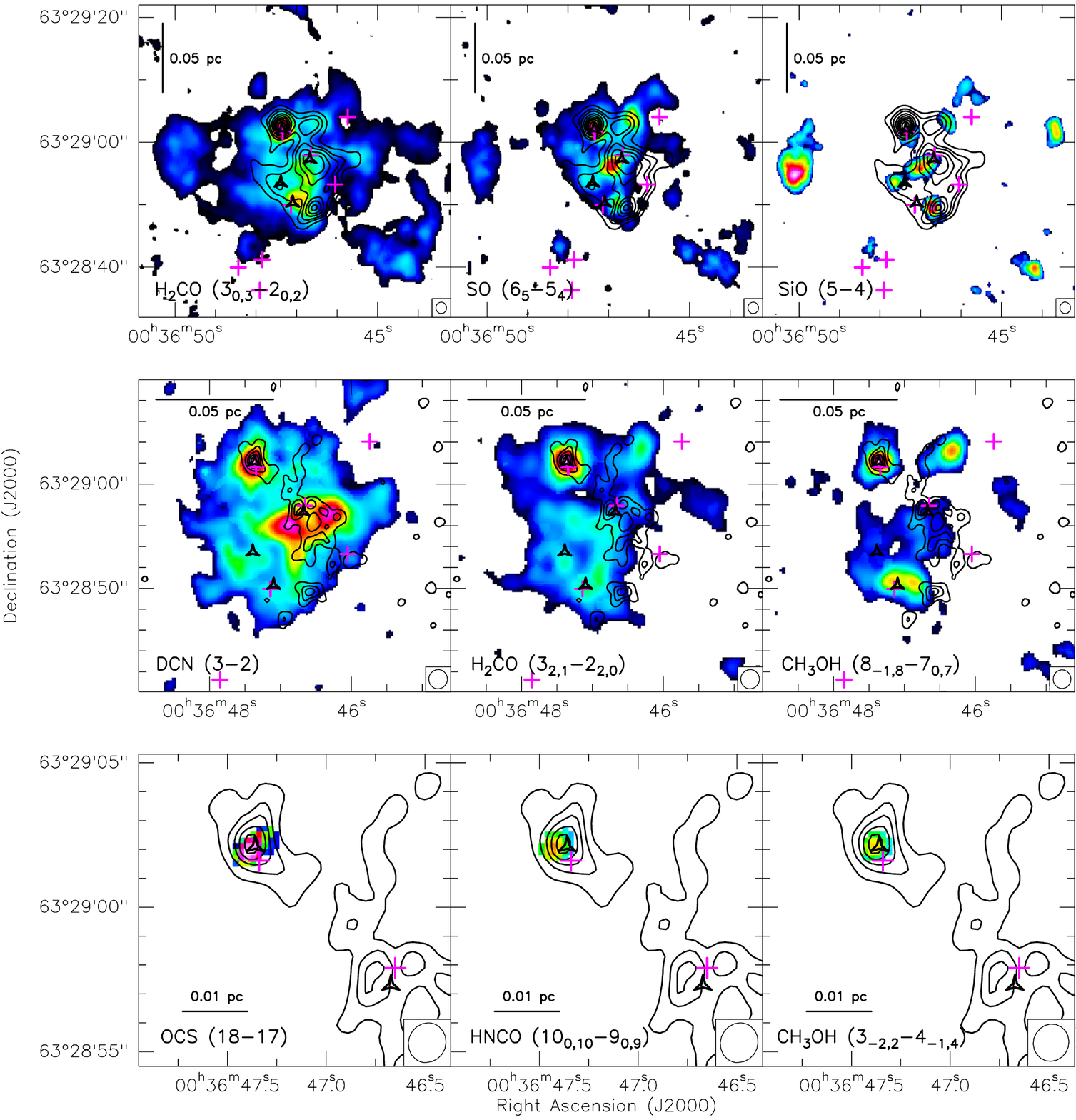}
\caption[Integrated intensity maps]{Color scale image of the integrated intensity (i.e., moment 0) maps of selected molecular line tracers overlapped with the contour map of the 1.3~mm dust continuum emission (only using the extended configuration for the dust continuum emission in the middle and lower panels). Pink crosses and black triangles indicate infrared \citep{Quanz07} and radio \citep{Anglada94} sources, respectively. The synthesized beam of each molecule is located at the bottom right corner.  % (only extended conf.)
}
\label{fig:moment0}
\end{figure*}
%

%%%%%%%%%%%%%%%%%%%%%%%%%%%%%%%%%%%%%%%%%%%%%%%%%%%%%%%%%%%%%%%%%%%%%%%%%%%%%%%%%%%%%%%%%%%%%%%%%%%%%
\section{Results} \label{sec:results}

%%%%%%%%%%%%%%%%%%%%%%%%%%%%%%%%%%%%%%%%%%%%%%%%%%%%%%%%%%%%%%%%%%%%%%%%%%%%%%%%%%%%%%%%%%%%%%%%%%%%%
\subsection{Dust continuum} \label{sec:continuum}

Figure \ref{Herschel-N_T} shows the large scale $N(H_{2})$ and $T_{d}$ maps of L1287.
On 1-10 pc scales this molecular cloud presents a filamentary morphology, with a dense molecular gas {\it hub} \citep[c.f.,][]{Myers2009} located at its central $\sim$0.5 pc region.
The $T_{d}$ map shows two higher temperature lobes to the northeast and southwest of the hub, which may be associated with heated cavities created by the bipolar CO outflow(s) emanated from the embedded low-mass YSO cluster \citep[][]{Snell90,Yang91,Quanz07}.

Figure~\ref{continuum} shows the SMA 1.3 mm continuum image on the central region of the hub.
The continuum emission at 1.3 mm can arise from dust thermal emission and free-free emission.
Based on the deep VLA observations of the centimeter free-free continuum emission presented in \citet{Anglada94}, we constrained the 1.3 mm free-free emission to be $<0.5$~mJy towards IRAS 0038+6312 and RNO\,1C.
When compared to the detected emission level in Figure~\ref{continuum}, we consider that the contribution of free-free emission is in general negligible. 

The 2$''$ resolution 1.3~mm dust continuum emission (top panel of Figure \ref{continuum}) splits up into six main cores (defined with at least a $35\sigma$ closed contour), which are spatially separated by a mean distance of $\sim0.03$~pc ($\sim6,500$~au), and have a mean size of $\sim0.02$~pc ($\sim4,500$~au). 
Assuming that the dust emission is optically thin, we estimated the gas and dust masses of these cores based on the following formulation
\begin{equation}\label{eq:thindust}
M=R\frac{d^2S_\nu}{B_\nu(T_d)\kappa_\nu},
\end{equation} 
where $d$ is the distance. 
Here we adopted $\kappa_\nu=0.9$~cm$^2$g$^{-1}$ for thin ice mantles after $10^5$ years of coagulation at a gas density of $10^6$~cm$^{-3}$ and for the frequency of our observations \citep{Ossenkopf94}. 
For these estimates we assumed $T_d$=22~K, which is the averaged dust temperature we derived in this region based on the Herschel data (Figure \ref{Herschel-N_T}, right).

The derived masses of these main six cores are in the range of $\sim0.4$--$4$~M$_\odot$.
The central core (SMA3; harboring RNO\,1C and VLA\,1) and the northeastern core (SMA1; harboring IRAS\,0038+6312 and VLA\,3) are the most massive ones, with $\sim4$ and $\sim2$~M$_\odot$, respectively. 
The faintest core SMA4, which may harbor VLA\,4, is located southeast of the central core, and has a mass of $\sim0.4$~M$_\odot$. 
The faint and small core (SMA5) located towards the south of the central core seems to be associated to the FU Orionis object RNO\,1B and has a mass of $\sim0.6$~M$_\odot$. 
The overall mass recovered by our SMA 1.3~mm continuum image is 22~M$_\odot$.

To better resolve the internal fragmentation in this region, we present the 1.3~mm continuum image generated using only the extended configuration data (i.e., $\sim$1$''$ angular resolution; Table \ref{observations}), which is shown in the bottom panel of Figure~\ref{continuum}.
The highest mass core SMA3 is resolved into a $\sim0.02$~pc scale clumpy toroid containing at least nine internal gas condensations; SMA3b and SMA3c may be associated with the YSO(s) RNO\,1C and VLA\,1.
SMA2 appears to be a $\sim0.02$~pc scale, elongated arm-like structure connecting to the clumpy toroid from the north; SMA5 and SMA6 may be parts of another $\sim0.04$~pc scale, elongated arm-like structure connecting to the clumpy toroid from the south.
% The core associated to IRAS 0038+6312 and VLA~3 (i.e., SMA1) does not seem to present further fragmentation, while SMA5, the core associated to RNO1B, has fragmented into two condensations. 
% The faintest cores SMA2 and SMA4 are not present in the high-angular resolution image probably due to lack of sensitivity or the filtering effect towards the more extended features. 
Overall, we identify 14 condensations with at least a $7\sigma$ closed contour. 
We measured the flux density of each condensation by fitting an elliptical Gaussian using the {\sc Miriad} task {\sc imfit}, and then based on Equation \ref{eq:thindust} we estimated the gas and dust masses. 
The polygons for fitting the Gaussian were drawn following the $6\sigma$ and the $9\sigma$ contour levels for the most extended sources ($1\sigma=0.5$ mJy~beam$^{-1}$). 
The derived masses range in between $\sim0.1$ and $\sim0.7$~$M_{\odot}$.
The properties derived from these fits also included the peak position, the deconvolved angular size and position angle and the intensity of the peak. The results are listed in Table~\ref{parameters}. 
Note that the total flux obtained taking into account only the extended configuration is $\sim16\%$ of the total flux obtained with all the configurations combined.
This is due to the filtering effect of the extended emission.

%
%\begin{landscape}
\begin{figure*}
\centering
\includegraphics[scale=0.8,keepaspectratio=true]{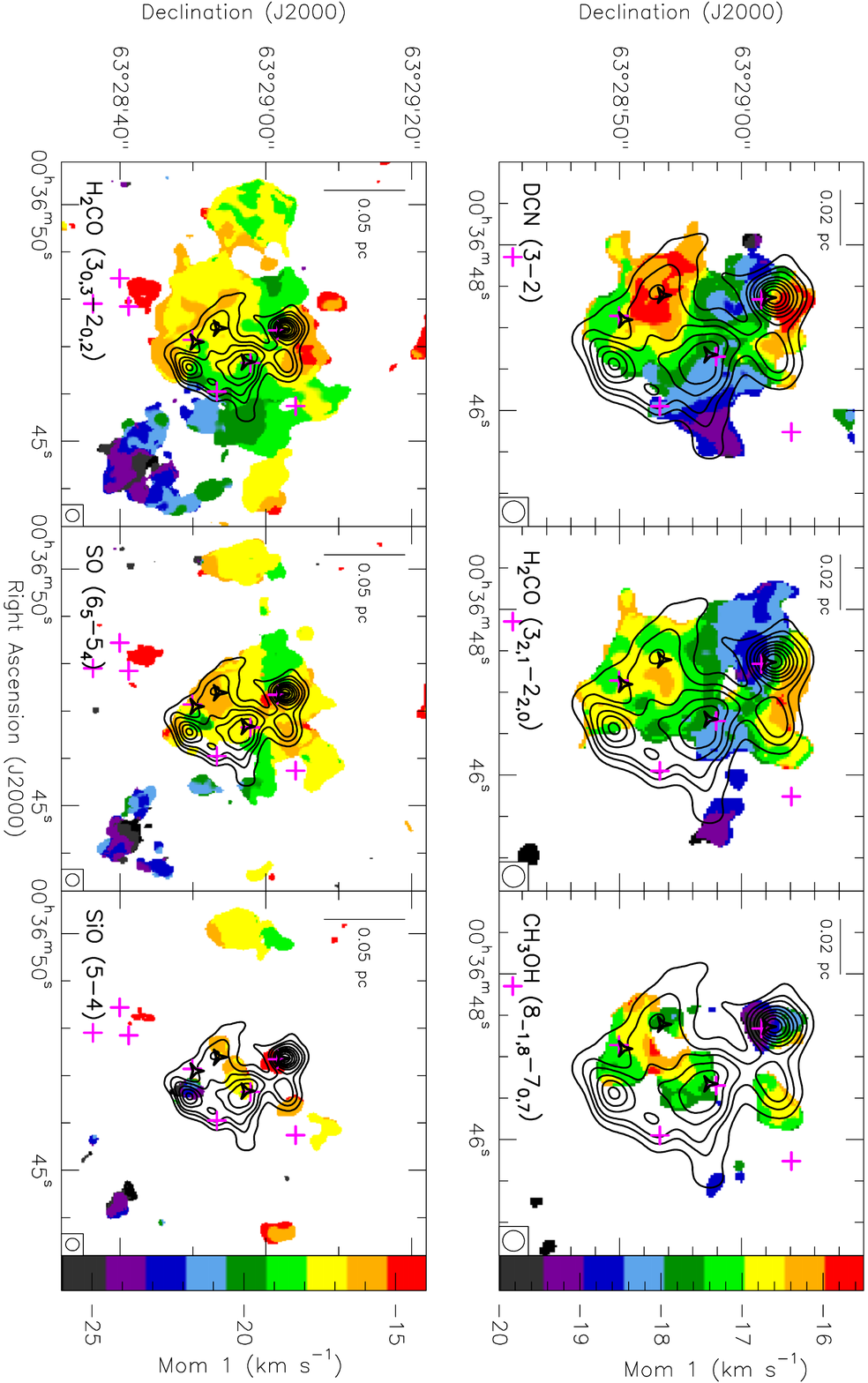}
\caption[Intensity-weighted averaged velocity maps]{Color scale image of the intensity-weighted averaged velocity (i.e., moment 1) maps of selected molecular line tracers overlapped with the contour map of the 1.3~mm dust continuum emission. Pink crosses and black triangles indicate infrared \citep{Quanz07} and radio \citep{Anglada94} sources, respectively. The synthesized beam of each molecule is located at the bottom right corner.  % (only extended conf.)
}
\label{fig:moment1}
\end{figure*}
%\end{landscape}
%
\begin{figure*}
\centering
\includegraphics[scale=0.55,keepaspectratio=true]{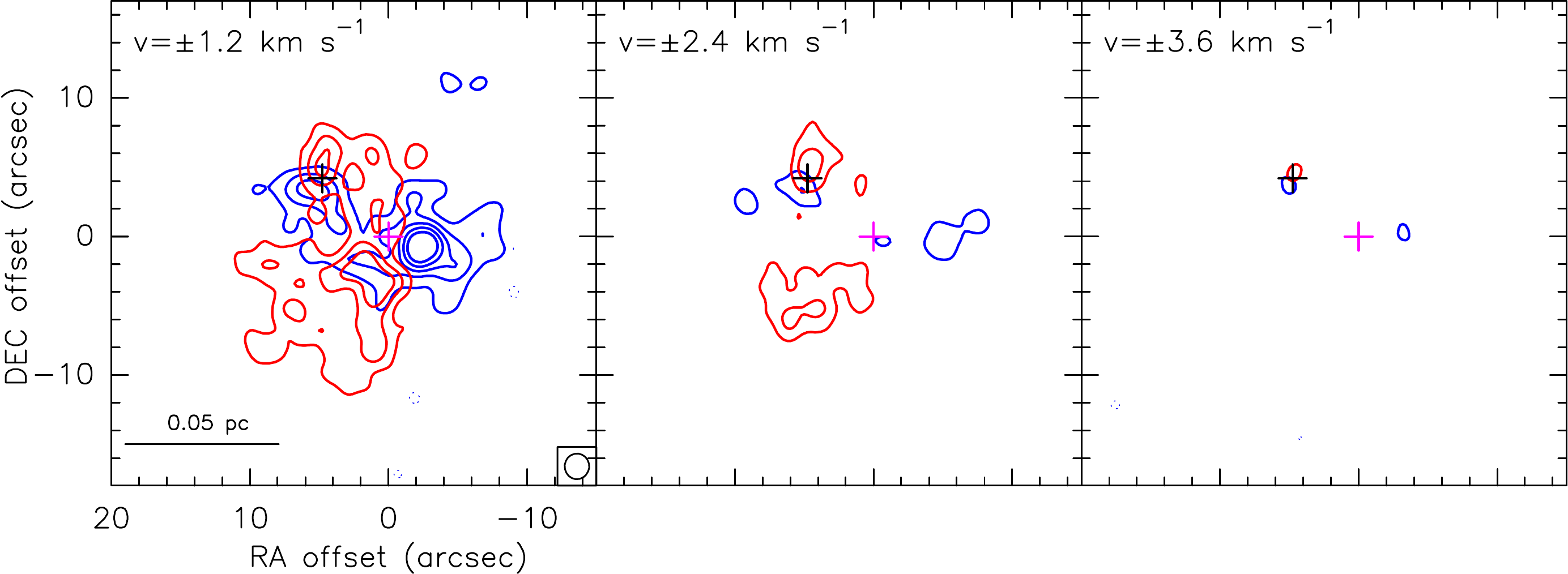}
\caption[DCN (3--2) channel map]{Each panel shows the blue- and red-shifted emission of DCN (3--2) at $\pm1.2$, $\pm2.4$ and $\pm3.6$~kms$^{-1}$ from the systemic velocity $-17.4$ km s$^{-1}$. Contours are $-4$, 4, 8, 12, ..., 20, 30, 40, 50 times the rms noise level of the map, 0.03~Jy~beam$^{-1}$. Black and pink crosses are IRAS 0038+6312 and RNO1C sources, respectively. The synthesized beam located at the bottom right corner of the first panel is $1\farcs85\times1\farcs77$, P.A. $=-6^\circ$.}
\label{fig:dcn-panels}
\end{figure*}
%

%\begin{landscape}
\begin{figure*}
 \centering
 \includegraphics[scale=0.65,keepaspectratio=true]{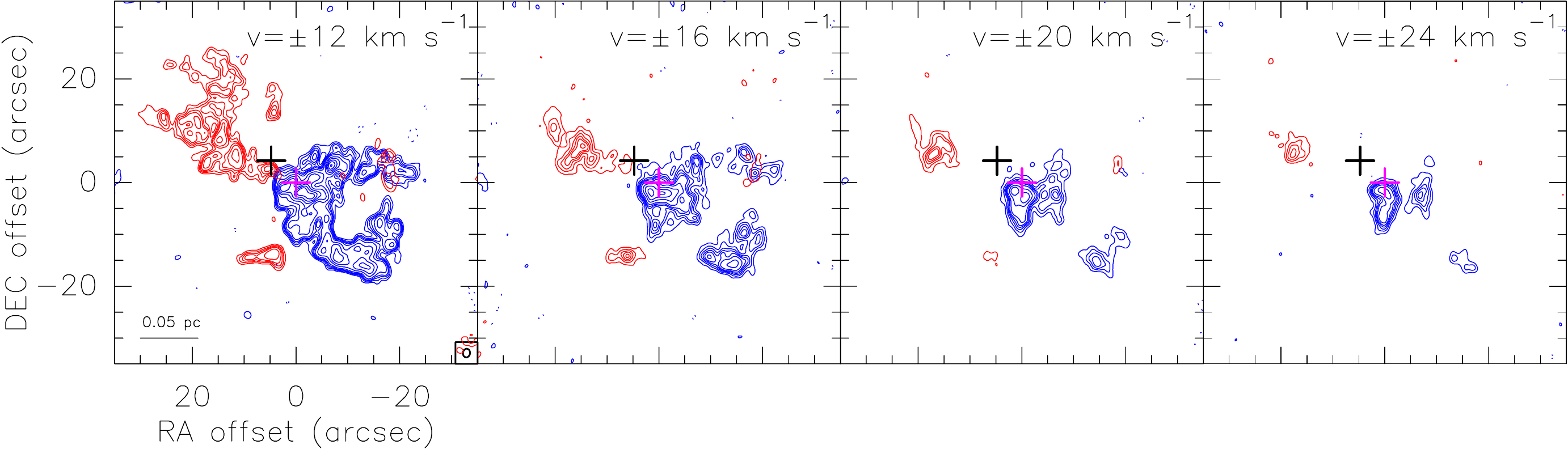}
  \caption[CO (2--1) high-velocity emission channel map]{CO (2--1) high-velocity emission. Each panel shows the red- and blue-shifted emission at $\pm12$, $\pm16$, $\pm20$ and $\pm24$ km s$^{-1}$ from the systemic velocity $-17.4$ km s$^{-1}$. Contour levels are: $-6$, 6, 9, ..., 15, 20, 30, 40, ..., 70 and $-3$, 3, 6, 9, ..., 15, 20, 30, 40, 50 times the rms level 0.04 Jy beam$^{-1}$ for the first and the rest of the panels, respectively. Black and pink crosses are IRAS 0038+6312 (also VLA~3) and RNO1C sources, respectively. The synthesized beam, located at bottom right of the first panel, is $1\farcs66\times1\farcs38$, P.A. $=-8^\circ$. A scale bar is located at the bottom left corner.}
  \label{COpanels}
\end{figure*}
%\end{landscape}
%
%
\begin{figure}[t]
\centering
\includegraphics[scale=0.35,keepaspectratio=true]{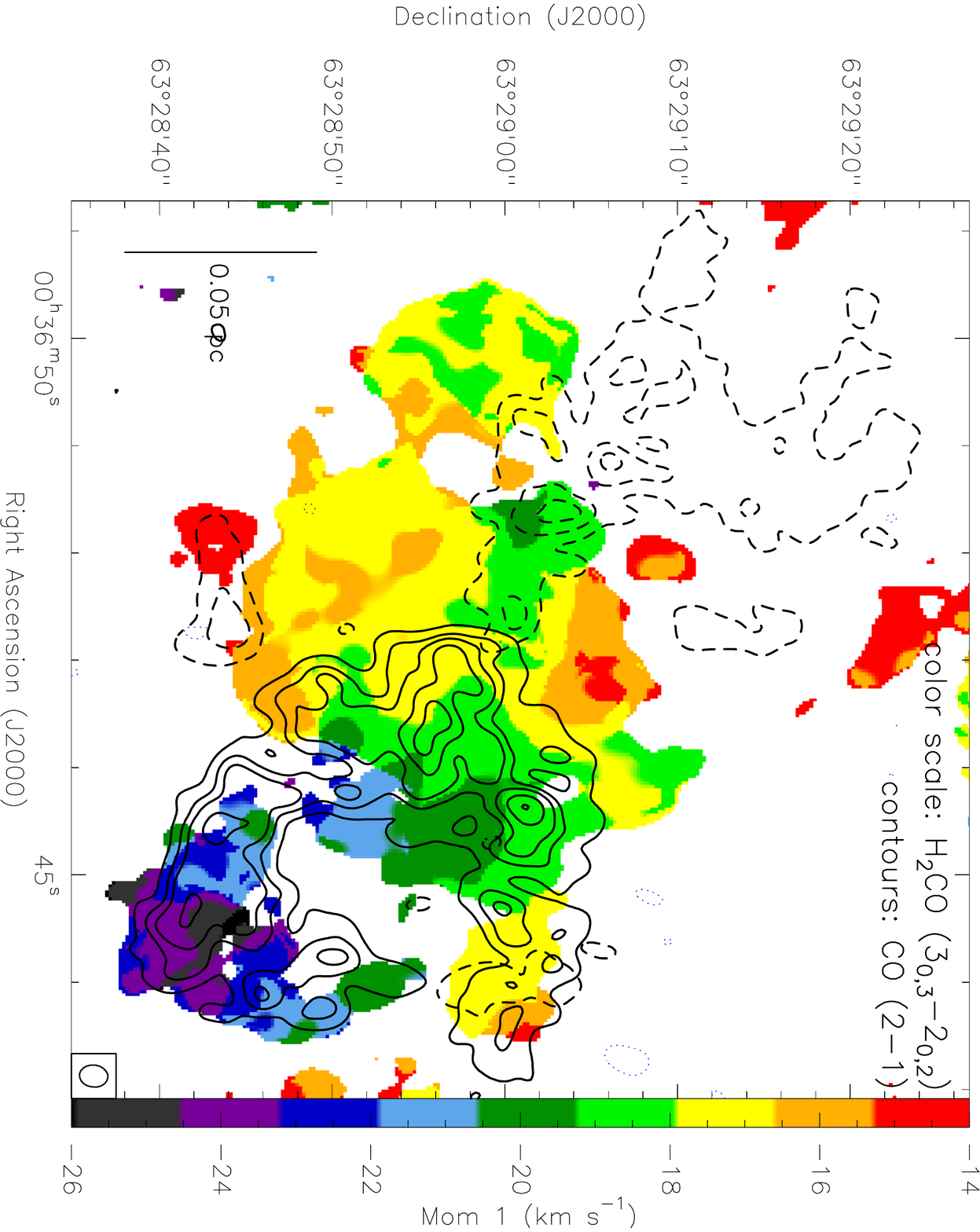}
\caption[H$_2$CO + CO emission]{{\it Contours}: CO (2--1) red- (dashed) and blue-shifted emission (solid) at $\pm12$~km~s$^{-1}$ from the systemic velocity $-17.4$~km~s$^{-1}$. Contour levels are: $-6$, 6, 18, 30, 50, 70 times the rms level 0.04~Jy~beam$^{-1}$. The synthesized beam located at the bottom right corner is  $1\farcs66\times1\farcs38$, P.A. $=-8^\circ$. {\it Color scale}: intensity-weighted averaged velocity (i.e., moment 1) map of H$_2$CO 3(0,3)--2(0,2) emission. A scale bar is located at the bottom left corner.
}
\label{fig:h2co+co}
\end{figure}
\begin{figure}[t]
 \centering
 \includegraphics[scale=0.55,keepaspectratio=true]{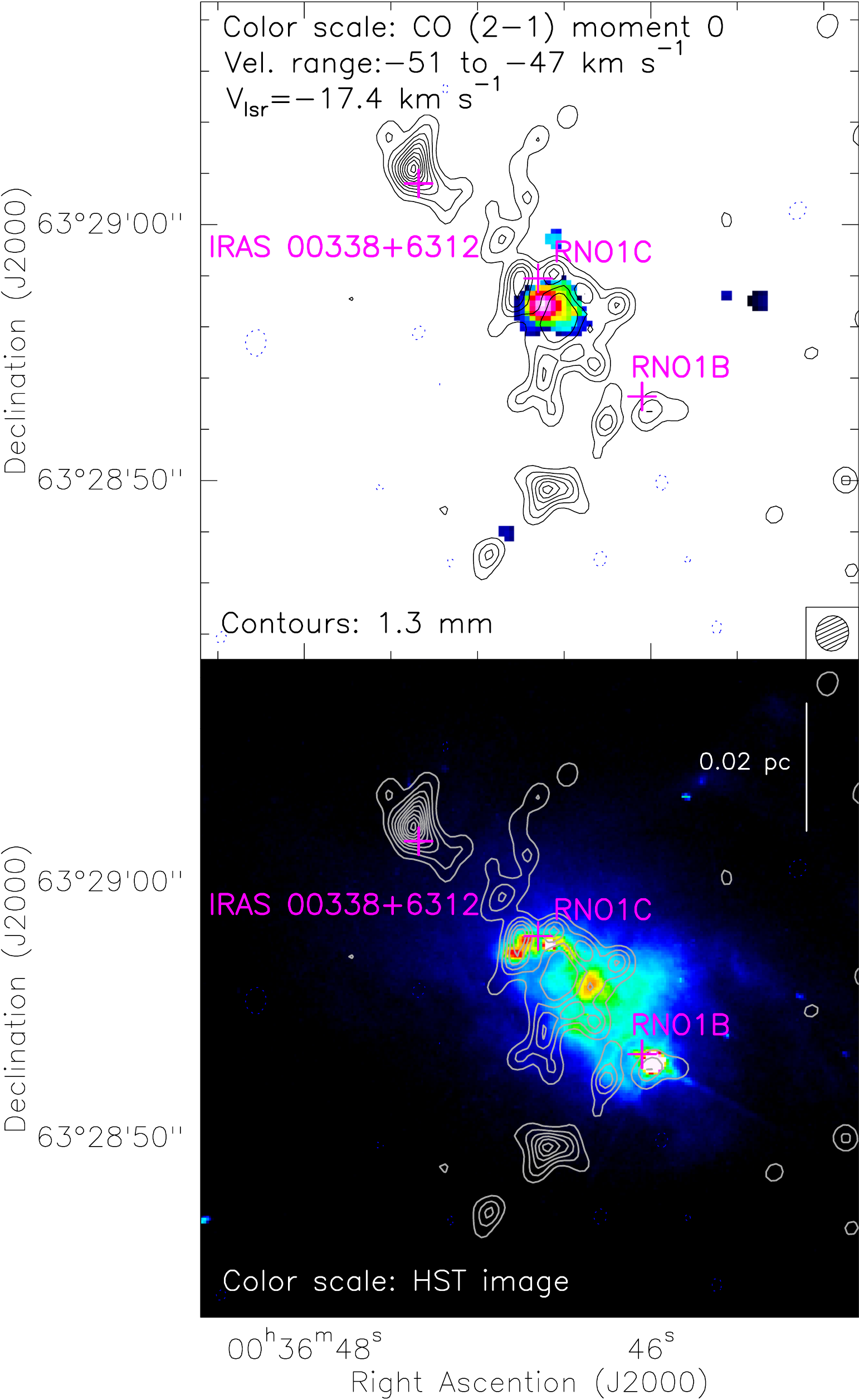}
  \caption{{\it Upper panel}: CO (2--1) highest-velocity emission. Color scale: Moment 0 (i.e., integrated intensity) map. Velocity range from $-51$ to $-47$ km s$^{-1}$. The synthesized beam located at the bottom right corner is $1\farcs39\times1\farcs27$, P.A. $=-19^\circ$. Contours: 1.3~mm dust continuum emission. Pink crosses indicate infrared sources \citep{Quanz07}. {\it Lower panel}: Color scale: Hubble Space Telescope image. WFPC2 instrument, F814W filter. %\textcolor{blue}{The lower panel is great!}
  }
  \label{CO21} %with extended and compact configurations data. Contour levels are 10, 30, 50, 70, 90 percent of the peak value, 14 mJy beam$^{-1}$.v$_{\text{lsr}}$ is $-17.6$ km s$^{-1}$
\end{figure}

%%%%%%%%%%%%%%%%%%%%%%%%%%%%%%%%%%%%%%%%%%%%%%%%%%%%%%%%%%%%%%%%%%%%%%%%%%%%%%%%%%%%%%%%%%%%%%%%%%%%%%
\subsection{Molecular lines} \label{sec:molecules}

In addition to the dust continuum emission, a number of molecular lines were detected with the SMA within the 8~GHz bandwidth. We identified 15 molecular lines of CH$_3$OH, SO, OCS, SiO, H$_2$CO, HNCO, DCN, DNC and CO. Table~\ref{transitions} of the Appendix lists the frequency, the energy of the upper level and the estimated critical density for each transition. In Fig.~\ref{fig:moment0}, we present the velocity integrated emission (i.e., moment 0) maps of several of the detected transitions in L1287.

The shock tracer SiO (5--4) reveals compact emission towards IRAS 00338+6312, RNO1C, VLA~4 and the dust continuum core SMA6. 
Also, it presents compact emission, neither associated to dust continuum emission nor to any YSO, towards the east, west and southwestern parts of the region. 
SO 6(5)--5(4) emission traces similar structures with those traced by SiO (5--4), but presents more prominent extended emission covering the central part of the observed region. 
In addition, the SO 6(5)--5(4) emission is more extended than SiO emission in the southwest, forming a ``V'' shaped structure.
CH$_3$OH 8($-1$,8)--7(0,7) and H$_2$CO 3(0,3)--2(0,2) and 3(2,1)--2(2,0) emission present similar morphology to SiO and SO, however, are additionally showing bright emission towards VLA~2 but are fainter around RNO\,1C. 
DCN (3--2) and H$_2$CO 3(2,1)--2(2,0) transitions trace well the dust continuum emission. 
The emission extends over the dust continuum presenting the strongest emission at IRAS\,0038+6312. 
In addition, DCN (3--2) shows strong emission towards the toroidal dust structure, with fainter emission coinciding with RNO\,1C and the center of the dust continuum toroid. 
Moreover, unresolved emission of CH$_3$OH 5(1,4)--4(2,2), CH$_3$OH 3($-2$,2)--4($-1$,4), OCS (18--17) and HNCO 10(0,10)--9(0,9) is detected at IRAS\,0038+6312.
% \\
%
%

%\subsection{Kinematics} \label{sec:kinematics}
In Figure~\ref{fig:moment1} we present the intensity-weighted averaged velocity (i.e., moment 1) maps of the lines presented in Figure \ref{fig:moment0}. 
A Gaussian fit to the DCN (3--2) spectrum averaged from the entire observed region 
indicates that the systemic velocity of this system is $-17.4$~km~s$^{-1}$. 
The dense core tracers H$_2$CO 3(0,3)--2(0,2) and 3(2,1)--2(2,0), SO 6(5)--5(4) and more clearly DCN (3--2) show a northwest-southeast large scale ($\sim0.1$~pc) velocity gradient between $-15$ and $-19$~km~s$^{-1}$ centered at the dust central core (SMA3) near RNO1C and VLA~1. 
We note that previous single dish observations of H$^{13}$CO$^{+}$ (1--0) also resolved a velocity gradient in the same direction, but is extended to a much larger angular scale \citep[$\sim$2$'$, $\sim$0.6 pc; see][]{Umemoto1999}.
In addition, on a smaller spatial scale and around IRAS\,0338+6312, these molecular lines also reveal a velocity gradient in the reversed direction.
These velocity gradients are perpendicular to the previously reported, northeast-southwest bipolar outflow  \citep{Snell90,Yang91}. 
The velocity gradient around IRAS\,0338+6312 is also seen in CH$_3$OH 8($-1$,8)--7(0,7). 
The emission towards the southwest of the dust continuum emission seen in  H$_2$CO, SO and SiO is clearly blue-shifted with a velocity of $\sim-24$~km~s$^{-1}$.

To study in more detail the kinematics in the region, we present the blue- and red-shifted emission of DCN (3--2) in Fig.~\ref{fig:dcn-panels}. 
The first panel presents the emission at $\pm1.2$~km~s$^{-1}$ from the systemic velocity. The blue- and red-shifted emission shows two structures forming the large-scale and reversed small-scale velocity gradients. 
At higher velocities (at $\pm2.4$ and $\pm3.6$~km~s$^{-1}$), the small-scale velocity gradient centered on IRAS 00338+6312 becomes more compact. 

%
%

%%%%%%%%%%%%%%%%%%%%%%%%%%%%%%%%%%%%%%%%%%%%%%%%%%%%%%%%%%%%%%%%%%%%%%%%%%%%%%%%%%%%%%%%%%%%%%%%%%%
\subsubsection{Molecular outflow} \label{sec:outflow}
In Figure~\ref{COpanels} we present the red- and blue-shifted high-velocity emission of the CO (2--1) in different velocity ranges. 
At these high velocities, the CO (2$-$1) traces molecular outflows. 
In agreement with the previous observations by  \citet{Snell90} and \citet{Yang91}, the emission shows an extended northeast-southwest bipolar molecular outflow centered near IRAS\,0038+6312 (and VLA\,3) and the FU Orionis object RNO\,1C. 
However, IRAS\,0038+6312 seems to fall closer to the center of the bipolar structure (see first and second panels of Fig.~\ref{COpanels}). 
Moreover, the red- and blue-shifted emission in the northeast and southwest lobes respectively, present clear cavity features, with the southern side of the blue-shifted lobe coinciding with the emission detected with the dense gas tracer H$_2$CO and the shock tracers SO and SiO. 
This strongly suggests that the outflow is dragging dense gas from this region (see Figures~\ref{fig:moment1} and~\ref{fig:h2co+co}), as it has already observed in other regions \citep[e.g. HH 2:][]{Girart05, Lefloch05}
In the last two panels, at the highest velocities, the blue-shifted emission appears at the position of RNO1C with an elongated structure in the north-south direction.  %at $\pm12$, $\pm16$, $\pm20$ and $\pm24$ km s$^{-1}$ from the systemic velocity $-17.4$ km s$^{-1}$

To analyze only the highest-velocity gas from the CO (2--1) emission, the upper panel of Figure~\ref{CO21} shows the integrated emission between $-51$ and $-47$ km s$^{-1}$. 
The emission at these velocities is compact and appears only very close to the position of RNO\,1C. 
It is also good to notice that the shock tracer SiO (5--4) shows emission towards both IRAS\,0038+6312 and RNO\,1C (see Figure~\ref{fig:moment0}).

\begin{figure}[t]
 \centering
\includegraphics[scale=0.8,keepaspectratio=true]{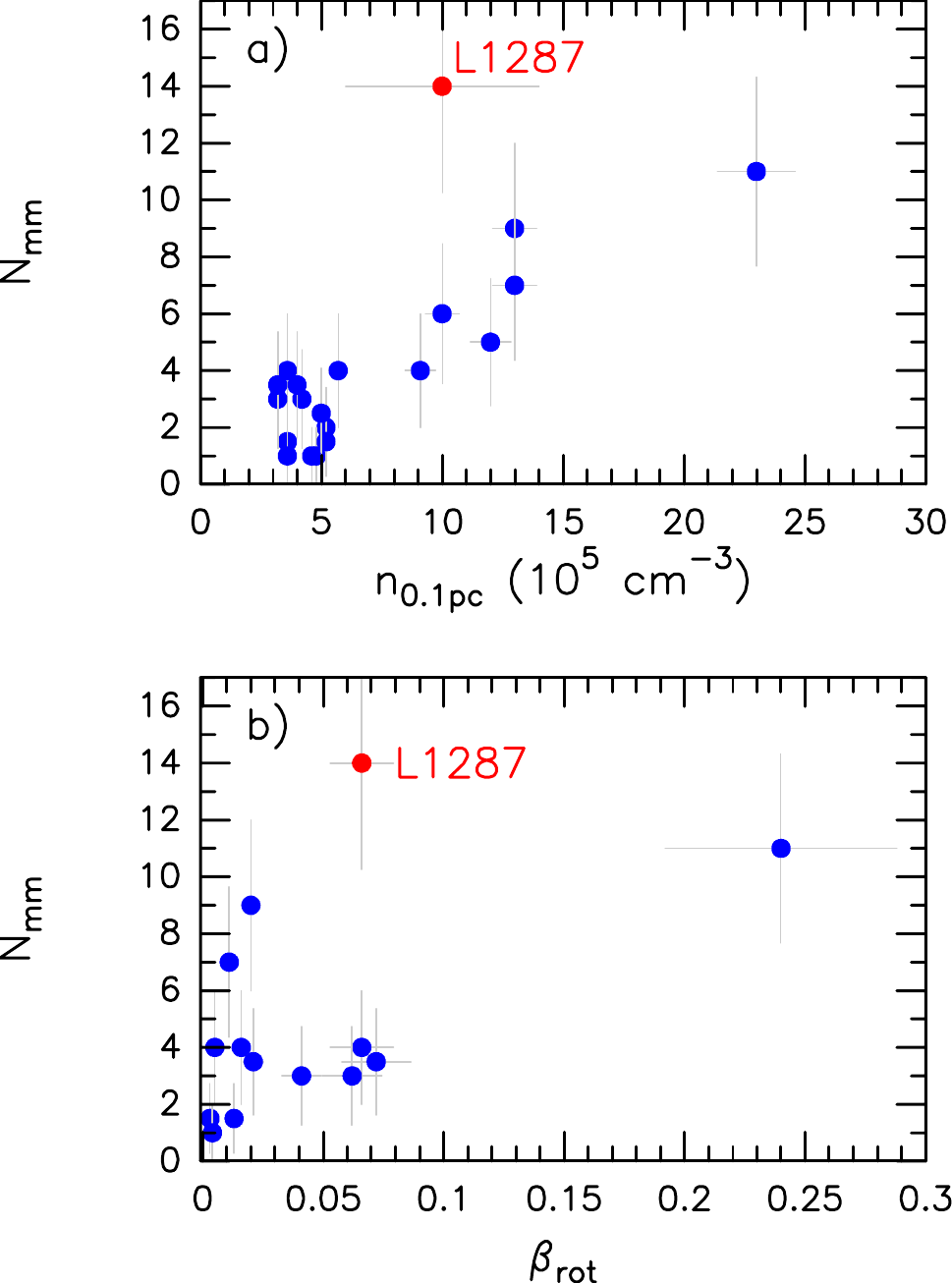}
\caption[]{
{\it Upper panel:} Fragmentation level ($N_\mathrm{mm}$) vs density averaged within a region of 0.1~pc of diameter after Palau et al. (2014, 2015). 
{\it Bottom panel:} fragmentation level vs rotational-to-gravitational energy after Palau et al. (2014). The location of L1287 in these plots is marked with a red symbol.
}
\label{fig:frag}
\end{figure}

%%%%%%%%%%%%%%%%%%%%%%%%%%%%%%%%%%%%%%%%%%%%%%%%%%%%%%%%%%%%%%%%%%%%%%%%%%%%%%%%%%%%%%%%%%%%%%%%%%%%%%%%%%%%%%%%
%%%%%%%%%%%%%%%%%%%%%%%%%%%%%%%%%%%%%%%%%%%%%%%%%%%%%%%%%%%%%%%%%%%%%%%%%%%%%%%%%%%%%%%%%%%%%%%%%%%%%%%%%%%%%%%%
\section{Discussion}\label{sec:analysis}
%%%%%%%%%%%%%%%%%%%%%%%%%%%%%%%%%%%%%%%%%%%%%%%%%%%%%%%%%%%%%%%%%%%%%%%%%%%%%%%%%%%%%%%%%%%%%%%%%%%%%%%%%%%%%%%%

\subsection{Fragmentation in L1287}
In Section 3.1 we presented the continuum emission at $1''$ angular resolution for L1287, which revealed 14 compact millimeter sources within a region of ~0.1 pc. Given the rms noise of the $1''$ image, our mass sensitivity is $\sim0.05$~M$_\odot$ (making the same assumptions as in Section 3.1). This mass sensitivity and our spatial resolution of 930~AU are well below the Jeans mass ($\sim0.6$~M$_\odot$) and Jeans length ($\sim6200$~AU) for a region of density $10^6$~cm$^{-3}$ and temperature of 20~K \citep[following][]{Palau15}, and are fully comparable to those reported by \citet{Palau14, Palau15}. Thus, the SMA observations presented here are well suited to study fragmentation in L1287.

\citet{Palau14, Palau15} study the fragmentation level of a sample of 19 intermediate and high-mass dense cores also within these spatial scale of 0.1 pc, and find fragmentation levels\footnote{The fragmentation level is defined by Palau et al. (2014, 2015) as the number of compact millimeter sources (with at least one closed contour) above 6 times the rms noise of the image, within a region of 0.1 pc of diameter.} ranging from 1 up to 11 sources. Thus, L1287, with 14 compact millimeter sources within 0.1 pc, presents a very high fragmentation level compared to the regions of this sample.

\citet{Palau14, Palau15} study the relation between the fragmentation level and different properties of the host cores, and find possible trends of fragmentation level increasing with the average density within 0.1 pc (with a correlation coefficient of 0.89) and with the ratio of rotational to gravitional energy (with a correlation coefficient of 0.57), usually referred to as $\beta_\mathrm{rot}$. Thus, we study here whether the average density within 0.1 pc and/or the rotation motions might explain the high fragmentation level in L1287.

In order to have a first approximation of the average density within 0.1 pc, we measured the flux density within an angular diameter of $22''$ (corresponding to 0.1 pc) in an image of the dust emission obtained with a single-dish, so that we avoid the filtering of large-scale emission produced by interferometers. We chose to use the James Clerk Maxwell Telescope data at 450~$\mu$m from \citet{DiFrancesco08} because this provides a main beam of $11''$ (and an effective beam of $17.3''$), which allows us to slightly resolve the emission within 0.1 pc. We obtained a flux density of 56 Jy within 0.1 pc for L1287. Taking into account an average dust temperature within 0.1 pc of 20 K (from the Herschel data presented in Section 2.2), this corresponds to an average density of around $10^6$~cm$^{-3}$. On the other hand, using the DCN velocity gradient of Fig. 10-bottom, we have estimated $\beta_\mathrm{rot}$ for L1287, which is of 0.067. In Fig.~\ref{fig:frag} we plot the fragmentation level vs average density within 0.1 pc and vs $\beta_\mathrm{rot}$ for the sample of \citet{Palau14, Palau15}, including the new results for L1287 (marked as a red symbol). These plots show that L1287 presents a rather high density and a high $\beta_\mathrm{rot}$, when compared to the entire sample of 19 massive dense cores of \citet{Palau14, Palau15}. Thus, it seems that both gravity and rotational energy could play an important role in the fragmentation process within L1287. 

Considering the compact nature of the millimeter fragments it seems reasonable to think that these fragments will form stars in the near future or are already associated with low-mass YSOs. Thus our data seem to indicate that L1287 is forming a dense cluster of deeply embedded low-mass YSOs.

\begin{figure}[t]
\centering
\includegraphics[scale=0.46,keepaspectratio=true,trim=10.5cm 0.2cm 10.0cm 0.9cm,clip]{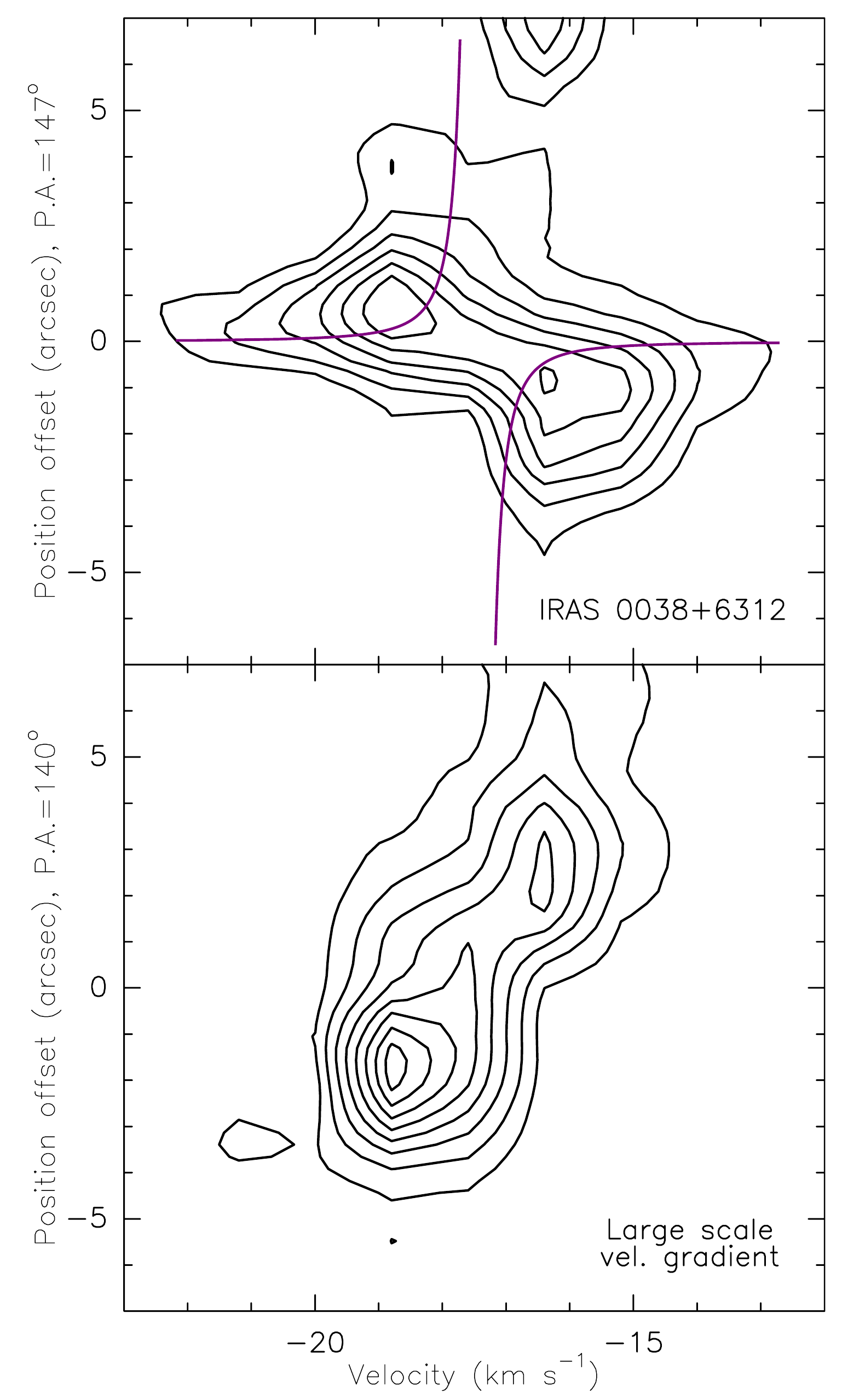}
\caption[DCN (3--2) position-velocity maps]{DCN (3--2) position-velocity maps. {\it Upper panel}: Emission along the velocity gradient at the IRAS 0038+6312 source. Contours are $-3$, 3, 5, 7, ..., 13 times the rms noise level of the map, 0.03 Jy beam$^{-1}$. The purple lines indicate a keplerian velocity distribution. {\it Lower panel}: Emission along the large velocity gradient passing through the center of the toroid traced by the dust continuum emission. Contours are $-3$, 3, 6, 9, ..., 24 times the rms noise level of the map, 0.03 Jy beam$^{-1}$.
}
\label{fig:dcn-pv}
\end{figure}
%

%\begin{figure}[t]
%

\begin{figure}[t]
 \centering
\begin{tabular}{ p{8.5cm} p{8.5cm} }
\includegraphics[scale=0.35,keepaspectratio=true,trim=6.3cm 1.2cm 6.2cm 4.9cm,clip]{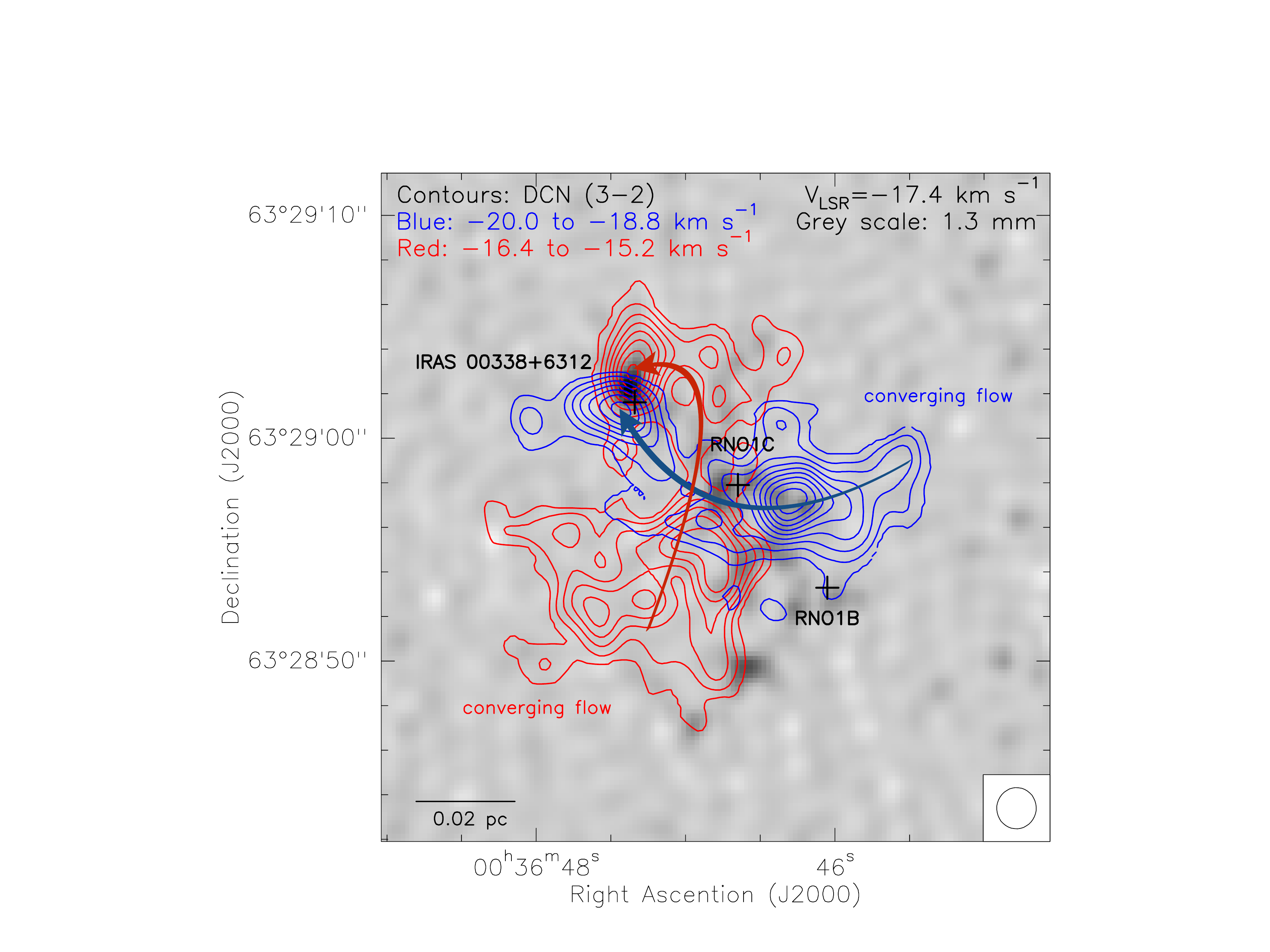}\\
\hspace{1.8cm}\vspace{-1cm} 
\includegraphics[scale=0.3,keepaspectratio=true]{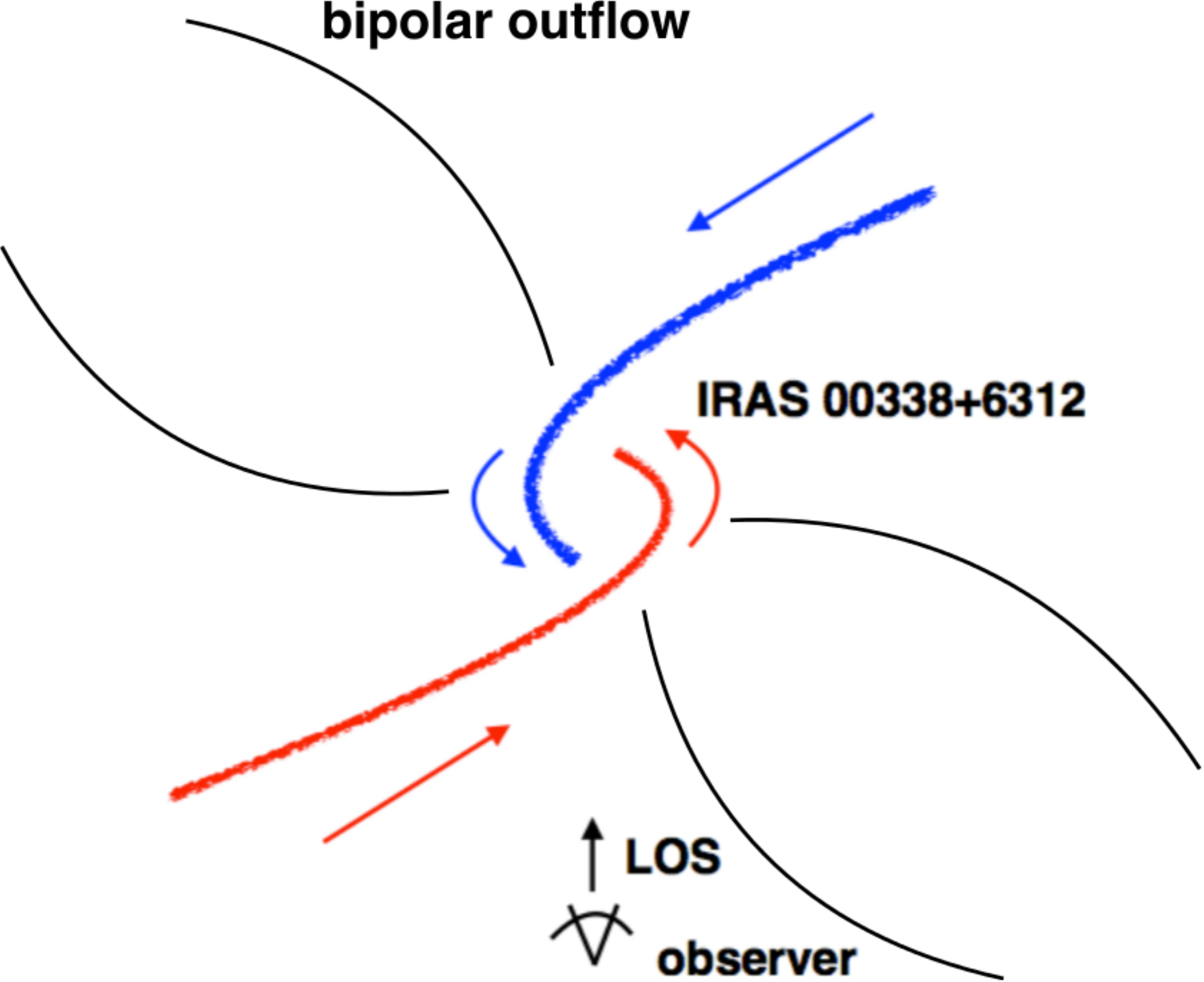}\\
\end{tabular}
\caption[Schematic model of the L1287 region. Converging flows+rotation dynamics]
{{\it Upper panel:} Contours correspond to the averaged emission from blue- and red-shifted channels of DCN (3--2) displaying the two low-velocity flow components. Contours are 6, 9, 12, 15, ..., 30 times the rms noise level of the map, 0.03 Jy beam$^{-1}$. The synthesized beam is located at the bottom right corner. The grey scale corresponds to the 1.3~mm dust continuum emission. The arrows indicate the two possible converging flows of the region. Black crosses indicate the position of infrared sources IRAS 00338+6312 and FU-Orionis RNO1B and RNO1C \citep{Quanz07}.
{\it Lower panel:} Schematic model of the L1287 region. Converging flows+rotation dynamics. The blue and red lines indicate the $\sim0.1$ pc scale molecular gas converging towards the center.  We can observe a reversed velocity gradient projected towards our line of sight as we can see at the IRAS 00338+6312 source.
}
\label{fig:diagram}
\end{figure}

\subsection{Dense gas kinematics}
For complicated cluster-forming regions resolved with CO outflows, it is always possible to attribute incoherent motions to the influence of outflow feedback or turbulence motions.
While with the data presented here we cannot rule out these cases, we discuss in the following one possibility to coherently interpret the observed dense gas kinematics from large to small spatial scales by gravitational contraction.

An inverse P Cygni line profile \citep{Pirogov16}, which is consistent with infalling motions on $\gtrsim$0.1 pc scales, has been found towards L1287 using single-dish observations of HCO$^{+}$ and H$^{13}$CO$^{+}$ (1--0).
On slightly larger scales, of $\sim$0.3-0.5 pc, \cite{Umemoto1999} resolved a blueshifted motion in the northwest, and a redshifted motion in the southeast.
On $\sim$0.1~pc scales, our DCN (3--2) and H$_{2}$CO 3(0,3)-2(0,2) line images (Figure \ref{fig:moment1}) trace a component of velocity gradient which has approximately the same orientation with the velocity gradient on $\sim$0.3-0.5 pc scales.
It is therefore natural to consider that our observations of these dense and/or warm gas tracers have revealed the innermost part of the coherent gas accretion flow.
\citet{Umemoto1999} interpreted both this large-scale velocity gradient by rotational motions, due to that it is perpendicular to the direction of the resolved CO outflow.
Alternatively, it remains possible to coherently interpret the large-scale ($\sim$0.1-0.5 pc) velocity gradient and the inverse P Cygni line profile by infall.
The presence of a CO outflow does not require a rotating disk on $\gtrsim$0.1 pc scales.
The velocities we detect from DCN (3--2) and H$_{2}$CO 3(0,3)-2(0,2) at $\sim$12$''$ (0.056 pc) from the center of SMA3 are $\sim$2~km\,s$^{-1}$.
The minimum required mass to gravitationally bind this motion is $\sim25$ M$_{\odot}$, which is consistent with the lower limit of the gas and dust mass we estimated based on the 1.3~mm dust continuum emission (22~M$_{\odot}$; see Sec.~\ref{sec:continuum}).

In this context, the reversed velocity gradient on smaller spatial scales around IRAS\,00338+6312 was unexpected.
In Figure~\ref{fig:dcn-pv}, we present the position-velocity (PV) diagrams made at the position angles of 140$^{\circ}$ and 147$^{\circ}$, %\textcolor{red}{X$^{\circ}$}
centered at SMA3 and IRAS\,00338+6312, respectively.
These two velocity gradients clearly have opposite directions.
Interestingly, the PV diagram made at the position of IRAS\,00338+6312 resembles that of a rotating disk, while the other PV diagram does not. From the IRAS\,0038+6312 velocity gradient, assuming Keplerian motion ($v=\sqrt{GM/R}$, where $v$ is the velocity, $G$ is the gravitational constant, $M$ is the mass and $R$ is the radius) and using a radius of $2.5''$ and $v=1.25$~km~s$^{-1}$, we estimated a mass for IRAS 0038+6312 of 4~M$_\odot$.  We note that the derived mass here is a lower limit as a possible inclination of the disk is not taken into account. 
The mass obtained is comparable to the $\sim6$~M$_\odot$ obtained by \citet{Yang91} using the mass-luminosity relation of (L/L$_\odot$)=(M/M$_\odot)^4$. 

If we consider that the small scale velocity gradient around IRAS\,00338+6312 is indeed due to rotational motion, then the reversion of the observed velocity gradients may be coherently interpreted as a filamentary large-scale inflow converging towards a circumstellar rotating disk, which may be further associated with the large-scale bipolar CO outflow (Section~\ref{sec:outflow}).
The schematic diagram presented in Figure \ref{fig:diagram} helps to illustrate this scenario. 
In the top panel of the figure, we present the blue- and red-shifted integrated DCN map
showing the two low-velocity bipolar structures with the reversed velocity gradients. In the bottom panel of the figure, we present our favored interpretation for the reversed velocity gradients. Taking into account the considerations given above, the reversed velocity gradients could be interpreted as converging flows towards the IRAS\,00338+6312 source.
The overall picture of forming a dense YSO cluster in this region from accreting dense gas filaments then may be analogous to those OB cluster-forming regions introduced in Section~\ref{sec:introduction}.

However, we have to point out that observationally it has been challenging to robustly distinguish rotational motion from infall, and is harder given the limited spectral resolution of our present observations.
While our present interpretation with converging flow was partly motivated by the $\gtrsim$0.1 pc scales inverse P Cygni line profile reported by \citep{Pirogov16}, we cannot yet rule out that our DCN data simply trace independent rotations around two centers, on smaller spatial scales.
The future observations of molecular lines with better spectral resolution and higher sensitivity, are still required to test our present interpretation.

%%%%%%%%%%%%%%%%%%%%%%%%%%%%%%%%%%%%%%%%%%%%%%%%%%%%%%%%%%%%%%%%%%%%%%%%%%%%%%%%%%%%%%%%%%%%%%%%%%%%%%%%%%%%%%%%
\subsection{Powering source of the northeast-southwest molecular outflow}
%\subsubsection{Powering source}
The powering source of the reported northeast-southwest bipolar outflow in L1287 has been a matter of debate in the past, as this is also related to how the accretion and outflows of YSOs are linked \citep[e.g.,][]{Evans1994}.
Several powering sources have been proposed, including RNO1B/1C \citep{Staude91} and VLA 3 or IRAS 0038+6312 \citep[e.g.,][]{Anglada94,Yang95,Quanz07}. 

By looking at the velocity range within $\pm$20\,km\,s$^{-1}$ with respect to the systemic velocity (Figure~\ref{COpanels}), the facts that IRAS\,00338+6312 is located closer to the center of the bipolar CO outflow (Section~\ref{sec:outflow}), and that the velocity gradient of the dense gas around it appears perpendicular to the direction of the outflow, indicate that IRAS\,00338+6312 is a probable powering source and may be associated with a circumstellar (pseudo-)disk.
However, at higher velocity (Figure~\ref{COpanels}) the bright and blueshifted CO emission appears more closely associated with RNO\,1C.
It may be possible that RNO\,1C is also powering a monopolar CO outflow.
The spatially compact nature of this outflow could be related to a recent accretion outburst event of this FU Orionis object.  %Probably there is no molecular gas which could interact with the red-shifted part of the outflow
A supporting evidence may be the Hubble Space Telescope $\sim8360$~\AA~(F814W filter) image shown in the lower panel of Figure~~\ref{CO21}.
At this wavelength the emission is associated with reflection or scattered light, and we would expect to detect it close to the blue-shifted part of the emission, which is the case. 
The brightest emission is clearly located towards the FU Orionis objects RNO\,1B and RNO\,1C. 
The emission also traces well the dust cavity structure seen with the high angular resolution image of the SMA with peaks of emission at the SMA3c and SMA3f millimeter sources.

%%%%%%%%%%%%%%%%%%%%%%%%%%%%%%%%%%%%%%%%%%%%%%%%%%%%%%%%%%%%%%%%%%%%%%%%%%%%%%%%%%%%%%%%%%%%%%%%%%%%%%%%%%%%%%%%
%%%%%%%%%%%%%%%%%%%%%%%%%%%%%%%%%%%%%%%%%%%%%%%%%%%%%%%%%%%%%%%%%%%%%%%%%%%%%%%%%%%%%%%%%%%%%%%%%%%%%%%%%%%%%%%%
\section{Conclusions} \label{sec:summary}
We have performed $\sim$1$''$ angular resolution SMA observations at 230 GHz towards the inner $\sim$0.25 pc cluster-forming region 
%condensed YSO cluster-forming region 
of the $\sim$10 pc scale filamentary dark cloud L1287.
Our 1.3~mm dust continuum image resolves a $\sim$0.02 pc scale clumpy dense gas toroid located at the center, which is surrounded by elongated, 0.02-0.04 pc scales arm-like dense gas structures and other spatially compact gas overdensities.
The fragmentation level found in the region, as compared to previous studies, is very high, with up to 14 compact millimeter sources within a region of 0.1~pc of diameter at a spatial resolution of $\sim1000$~AU.

In addition, our observations of the dense molecular gas tracers DCN (3--2) and H$_{2}$CO 3(0,3)-2(0,2) resolved two components of velocity gradients: one has approximately the same direction as the velocity gradient traced by previous H$^{13}$CO$^{+}$ observations on 0.3-0.5 pc scales; the other one is seen on smaller spatial scales around the embedded YSO IRAS 00338+6312, and presents a reversed direction.
We cannot yet rule out that such incoherent motions are due to the influence of outflow feedback, or due to turbulent motions; and cannot yet rule out that we are seeing independent rotational motions around two centers.
However, we found that these motions might be coherently interpreted as a filamentary, 0.1-0.5 pc scale infalling gas flow, which is converging towards the rotating circumstellar disks on smaller spatial scales.
The formation of the low-mass YSO cluster at the center of L1287 might be fed by the global gas inflow, which is analogous to some luminous OB cluster-forming regions such as W49A, G10.6-0.4, G33.92+0.11, or NGC6334\,V.

%%%%%%%%%%%%%%%%%%%%%%%%%%%%%%%%%%%%%%%%%%%%%%%%%%%%%%%%%%%%%%%%%%%%%%%%%%%%%%%%%%%%%%%%%%%%%%%%%%%%%%%%%%%%%%%%
\begin{acknowledgements}
We thank the SMA staff for their support which makes these studies possible. CJ acknowledges support from MINECO (Spain) BES-2012-052481 grant. CJ, JMG and GB acknowledge support from MICINN (Spain) AYA2014-57369-C3 and AYA2017-84390-C2-2-R grants. 
AP acknowledges financial support from UNAM and CONACyT, M\'exico.
RGM acknowledges support from UNAM-PAPIIT program IA102817.
\end{acknowledgements}

\begin{appendix}
\section{Line identifications}

In this Appendix we present a table with the list of detected transitions towards L1287 with the SMA (Table~\ref{transitions}).

\begin{table}[h]
\caption{Molecular lines detected with the SMA towards L1287}
\begin{center}
\begin{tabular}{c c c c c}
\hline
\hline
Molecule		&Transition				&Frequency	&E$_{\text{U}}$		&$\rho_{\text{crit}}^\mathrm{a}$\\ 
  				&			 			&(GHz)		&(K) 			  	&($\text{cm}$$^{-3}$)\\ 
\hline
CH$_3$OH		&5(1,4)--4(2,2)			&216.946	&56				 	&2.33$\times10^{6}$\\
SiO				&(5--4)					&217.105	&31				 	&3.25$\times10^{8}$\\
DCN				&(3--2)					&217.239	&21					&--\\
H$_2$CO			&3(0,3)--2(0,2)			&218.222	&21					&2.56$\times10^{6}$\\
H$_2$CO			&3(2,2)--2(2,1)			&218.476	&68				 	&2.96$\times10^{6}$\\
H$_2$CO			&3(2,1)--2(2,0)			&218.760	&68					&3.36$\times10^{6}$\\
OCS				&(18--17)				&218.903	&100				&4.28$\times10^{5}$\\
C$^{18}$O		&(2--1)					&219.560	&16					&9.33$\times10^{3}$\\
HNCO			&10(0,10)--9(0,9)		&219.798	&58					&--\\
SO				&6(5)--5(4)				&219.949	&35					&--\\
$^{13}$CO		&(2--1)					&220.399	&16					&9.38$\times10^{3}$\\
DNC				&(3--2)					&228.910	&22					&--\\
CH$_3$OH		&8($-1$,8)--7(0,7)		&229.759	&89					&2.62$\times10^{7}$\\
CH$_3$OH		&3($-2$,2)--4($-1$,4)	&230.027	&40					&7.05$\times10^{6}$\\
CO				&(2--1)					&230.538	&17					&1.07$\times10^{4}$\\
\hline	
\end{tabular}
\end{center}
\label{transitions}
$^\mathrm{a}$ $\rho_{\text{crit}}$=A$_{\text{ul}}/\gamma$, where A$_{\text{ul}}$ is the Einstein spontaneous emission coefficient and $\gamma$=$\sigma<$v$>$ is the collisional rate where $\sigma$ is the cross section of the collision for each transition and $<$v$>\approx$(3kT/m)$^{1/2}$ is the average velocity of the collisional particles. As H$_2$ is the most abundant molecule it is used as the dominant collisional particle. The Einstein spontaneous emission coefficients are taken from LAMDA$^\mathrm{b}$ and splatalogue.net databases and $\gamma$ values are taken from LAMDA database. We used T=20~K (as 22~K is not available in the tabulated data). \\  
$^\mathrm{b}$ http://home.strw.leidenuniv.nl/$\sim$moldata/
\end{table}

\end{appendix}

   \bibliographystyle{aa} % style aa.bst
   \bibliography{bibliography} % your references Yourfile.bib

\end{document}